\documentclass[fleqn,usenatbib, referee]{mnras}

\usepackage{newtxtext,newtxmath}
\usepackage{subfig}
\usepackage{float}
\usepackage{footnote}
\makesavenoteenv{tabular}

\usepackage[T1]{fontenc}
\usepackage{ae,aecompl}
\usepackage{gensymb}

\usepackage{graphicx}	
\usepackage{amsmath}	
\usepackage{amssymb}	

\title[Pulse Frequency Fluctuations of Magnetars]{Pulse Frequency Fluctuations of Magnetars}

\author[D. \c{C}erri--Serim, M. M. Serim, \c{S}. \c{S}ahiner, S. \c{C}. \.{I}nam and A. Baykal]
{D. \c{C}erri--Serim$^{1}$\thanks{E-mail: danjela@astroa.physics.metu.edu.tr (D\c{C}S); muhammed@astroa.physics.metu.edu.tr (MMS); seyda@astroa.physics.metu.edu.tr (\c{S}\c{S}); inam@baskent.edu.tr (S\c{C}\.{I}); altan@astroa.physics.metu.edu.tr (AB)}, M. M. Serim$^{1}$\footnotemark[1], \c{S}. \c{S}ahiner$^{1}$\footnotemark[1], S. \c{C}. \.{I}nam$^{2}$\footnotemark[1] and A. Baykal$^{1}$\footnotemark[1] \\
$^{1}$Physics Department, Middle East Technical University, 06531 Ankara, Turkey\\
$^{2}$Department of Electrical and Electronics Engineering, Ba\c{s}kent University, 06790 Ankara, Turkey}

\date{Accepted XXX. Received YYY; in original form ZZZ}

\pubyear{2018}

\begin{document}
\label{firstpage}
\pagerange{\pageref{firstpage}--\pageref{lastpage}}
\maketitle


\begin{abstract}

Using \emph{RXTE}, \emph{Chandra}, \emph{XMM-Newton} and \emph{Swift} observations, we for the first time construct the power spectra and torque noise strengths of magnetars. For some of the sources, we measure strong red noise on timescales months to years which might be a consequence of their outbursts. We compare noise strengths of magnetars with those of radio pulsars by investigating possible correlations of noise strengths with spin-down rate, magnetic field and age. Using these correlations, we find that magnetar noise strengths are obeying similar trends with radio pulsars. On the contrary, we do not find any correlation between noise strength and X-ray luminosity which was seen in accretion powered pulsars. Our findings suggest that the noise behaviour of magnetars resembles that of radio pulsars but they possess higher noise levels likely due to their stronger magnetic fields.

\end{abstract}


\begin{keywords}
stars: neutron - pulsars: general - stars: magnetars
\end{keywords}



\section{Introduction}

Magnetars are isolated young pulsars which have extremely strong inferred dipolar magnetic fields; the strongest among astronomical objects known to date. X-ray observations of magnetars have revealed several interesting observational phenomena including giant flares, large outbursts, short bursts, and quasi periodic oscillations. Magnetars have also shown to exhibit distinct timing properties such as enhanced spin-down, glitches and anti-glitches \citep[see][for a comprehensive review]{kaspi2017}. An accreting system, SXP 1062 have also shown a glitch, with a magnetar field deduced from its spin-down rate \citep{serim2017b}. Using pulse timing techniques, it is possible to figure out glitches and anti-glitches, and study the characteristics of timing noise of magnetars. Glitch and anti-glitch are the sudden jumps observed in the pulse frequency time series of pulsars; whereas timing noise, being related to stochastic variations in time series, is a measure of irregularities in the pulse frequency time series. Timing noise was discussed firstly for Crab pulsar \citep{boynton1972}, and it was found that its noise behaviour corresponds to random walk in frequency. 

Theoretical studies have shown that power density spectra of red-noise in timing residuals are generally frequency dependent and proportional to  $\sim$ $f^{\alpha}$ ,where $\alpha=-1, -3, -5$ corresponds to the timing noise in phase, frequency and spin-down rate, respectively \citep[e.g][]{dal1995}. Estimating power density spectra of timing noise is not straightforward because; in general, the data are sampled unevenly in long time scales ranging from days to years. The idea that timing noise is a kind of a random walk process is still valid to date, however, this model is too simple and idealized to explain the noise phenomenon. A detailed model describing timing noise was presented by \cite{cordes1985} where they showed that random walk processes were the result of superimposing of some discrete micro-jumps in the timing parameters of the source. Throughout the years it has been shown that timing noise is weakly correlated with period and strongly correlated with period-derivative (implying a strong anti-correlation with characteristic age) \citep{cordeshelfand1980,dal1995,hobbs2010}.

Until 2010, there were only a few published studies about calculation of timing noise of pulsars from longer time spans of about $>10$ years \citep{shaban1995,baykal1999,stairs2000,shaban2001}. In 2010, \cite{hobbs2010} analysed red-noise in timing residuals of 366 pulsars with a time span which can go up to 34 years, concluding that timing noise is weakly correlated with magnetic field and that timing residuals show a quasi-periodic pattern on long timescales. In addition, they suggested that timing noise of pulsars with a characteristic age $<10^5$ yr can be the result of glitch recoveries and that 
the process causing slow-glitches (i.e. an increase in frequency but stable spin-down rate) should not be different than that of timing noise.

In this paper, we present our calculations of the timing noise of all magnetars using the available data up-to-date. To reach our aim, we either re-construct a given timing solution by phase-connecting the time of arrivals, or we have used the frequency histories found in the literature (see Table \ref{sources} for references). In order to avoid excess noise appearing due to the pulse frequency changes during the glitch events, we exclude time intervals with reported glitch events from light curves in the construction of power density estimation. Out of twenty-three confirmed magnetars, two are not studied because of the lack of data and noise measurements of another three magnetars were already performed in our previous work \citep{serim2012}. For the remaining eighteen sources we either measure the noise strength or we estimate the power density spectra depending on the
sampling rate of the data. The noise strength values are then compared with the noise estimates of radio pulsars. This paper advances as follows. In Section 2, we describe data reduction procedures. In Section 3, the methodology of noise analysis and power density spectra estimation are explained. The results are reported and discussed in Section 4. 

\section{Data Reduction}

For this work, we perform the analysis of \textit{XMM-Newton}, \textit{Chandra}, \textit{Swift} and \textit{RXTE} observations of eighteen magnetars. We select the data of glitch-free time intervals for our analysis. All data are gathered from NASA's HEASARC Archive.

For \textit{XMM-Newton} observations, data reduction is carried through \verb"SAS v.15.0.0" software. For the observations with high energy background, we exclude the time intervals in which background level is higher than 5 per cent of the source flux. Data are filtered for event patterns 0--4 with only good events (FLAG=0). Source events for all observations are extracted from a 20 arcsec circle centred on the source coordinates. Background events are obtained from a circular source-free region on the same CCD.

\textit{Chandra} data products are examined via \verb"CIAO v.4.9" software. The data is reprocessed to produce the level 2 event files using \verb"REPRO" tool with \verb"CALDB v.4.7.6". A 5 arcsec circle centred on source coordinates is used to gather source counts, while a source-free 10 arcsec circle is used to extract background events on the same CCD.

Data reduction of \textit{RXTE}--PCA data is deployed with \verb"HEASOFT v.6.19" software. Data are filtered with the options of electron contamination to be less than 0.1, elevation angle to be greater than 10 and offset angle less than 0.02. From these cleaned event files, light curves with a resolution of 0.05 s are created. The light curves are corrected for PCU on/off status using \verb"CORRECTLC" tool within \verb"Ftools" software. 

\textit{Swift}--XRT data are cleaned with the standart pipeline script \verb"XRTPIPELINE V.0.13.2" using default screening criteria. Clean event files are barycentered using source coordinates and up-to-date clockfile. Light curves are extracted with \verb"XSELECT V.2.4D". For windowed timing mode observations, no region selection is applied during light curve extraction. For photon counting mode observations, a 20 pixel circular source region and a 60 pixel circular background region are used. 

\section{Noise and Power Density Spectra Estimation}

\begin{table*}
  \caption{List of magnetars and their properties.}
  \label{sources}
  \center{\renewcommand{\arraystretch}{1.25}\begin{tabular}{@{}l@{}cc@{}ccl@{}c@{}c@{}}
  \hline
 Source name & $P$ $^{(a)}$ &  $\dot{P}$ $^{(a)}$ & $B_{\mathrm{surf}}$ $^{(a)}$ & $L_{\mathrm{x}}$ $^{(a)}$ & $\log(S)$ $^{(b)}$ & Time Span $^{(b)}$ & Timing Parm. \\
  & (s) & ($10^{−11}$ s/s) & ($10^{14}$ G) & ($10^{33}$ erg/s) & ($\log$ (s$^{-3}$)) & (d) & References \\
 \hline 
CXOU J010043.1--721134	&	8.020392(9)	&	1.88(8)	&	3.9	&	65	&	-20.47$^{+1.19}_{-0.44}$	&	1881	&	1, 2 $^{(c)}$	\\
4U 0142+61	&	8.68869249(5)	&	0.2022(4)	&	1.3	&	105	&	-22.65$^{+1.19}_{-0.44}$	$^{**}$ &	993	&	3 $^{(d)}$ (Eph B,C,D) 	\\
SGR 0418+5729	&	9.07838822(5)	&	0.0004(1)	&	0.061	&	0.00096	&	-23.68$^{+1.19}_{-0.44}$	&	1865	&	4 $^{(d)}$	\\
SGR 0501+4516	&	5.7620695(1)	&	0.594(2)	&	1.9	&	0.81	&	-21.63$^{+1.19}_{-0.44}$	&	547	&	5 $^{(d)}$	\\
SGR 0526--66	&	8.0544(2)	&	3.8(1)	&	5.6	&	189	&	-16.91$^{+1.19}_{-0.44}$	&	3543	&	6, 7 $^{(c)}$	\\
1E 1048.1--5937	&	6.457875(3)	&	2.25	&	3.9	&	49	&	-18.54$^{+0.27}_{-0.18}$	&	972	&	3 $^{(e)}$	\\
1E 1547.0--5408	&	2.0721255(1)	&	4.77	&	3.2	&	1.3	&	-15.56$^{+1.19}_{-0.44}$	&	810	&	8 $^{(f)}$	\\
PSR J1622--4950$^{(g)}$	&	4.3261(1)	&	1.7(1)	&	2.7	&	0.44	&	--	&	--	&	--	\\
SGR 1627--41$^{(g)}$	&	2.594578(6)	&	1.9(4)	&	2.2	&	3.6	&	--	&	--	&	--	\\
CXOU J164710.2--455216	&	10.610644(17)	&	<0.04	&	<0.66	&	0.45	&	-21.66$^{+1.19}_{-0.44}$	&	1066	&	9 $^{(d)}$	\\
1RXS J170849.0--400910	&	11.00502461(17)	&	1.9455(13)	&	4.7	&	42	&	-21.71$^{+1.19}_{-0.44}$	$^{**}$ &	972	&	3 $^{(d)}$ (Eph C,F)	\\
CXOU J171405.7--381031	&	3.825352(4)	&	6.40(5)	&	5	&	56	&	-16.63$^{+1.19}_{-0.44}$	&	370	&	10, 11 $^{(c)}$	\\
SGR J1745--2900	&	3.76363824(13)	&	1.385(15)	&	2.3	&	<0.11	&	-17.70$^{+1.19}_{-0.44}$	&	494	&	12 $^{(d)}$	\\
SGR 1806--20	&	7.54773(2)	&	49.5	&	20	&	163	&	-17.33$^{+0.27}_{-0.18}$	&	1063	&	13 ... 22 $^{(c)}$	\\
XTE J1810--197	&	5.5403537(2)	&	0.777(3)	&	2.1	&	0.043	&	-19.87$^{+0.27}_{-0.18}$	&	878	&	23 ... 31 $^{(c)}$	\\
Swift J1822.3--1606	&	8.43772106(6)	&	0.0021(2)	&	0.14	&	<0.00040	&	-22.60$^{+1.19}_{-0.44}$	$^{*}$ &	842	&	32 $^{(d)}$	\\
SGR 1833--0832	&	7.5654084(4)	&	0.35(3)	&	1.6	&	--	&	-20.52$^{+1.19}_{-0.44}$	$^{*}$ &	275	&	33 $^{(d)}$	\\
Swift J1834.9--0846	&	2.4823018(1)	&	0.796(12)	&	1.4	&	<0.0084	&	-19.60$^{+1.19}_{-0.44}$	$^{*}$ &	100	&	34 $^{(d)}$	\\
1E 1841--045	&	11.788978(1)	&	4.092(15)	&	7	&	184	&	-21.06$^{+0.27}_{-0.18}$	$^{**}$ &	1016	&	3 $^{(d)}$ (Eph A,C,E)	\\
3XMM J185246.6+003317	&	11.55871346(6)	&	<0.014	&	<0.41	&	<0.0060	&	-23.62$^{+1.19}_{-0.44}$	&	215	&	35, 36 $^{(c)}$	\\
SGR 1900+14	&	5.19987(7)	&	9.2(4)	&	7	&	90	&	-18.19$^{+1.19}_{-0.44}$	&	1058	&	15, 22,  $^{(c)}$	\\
SGR 1935+2154	&	3.2450650(1)	&	1.43(1)	&	2.2	&	--	&	-20.64$^{+1.19}_{-0.44}$	&	284	&	37  $^{(d)}$	\\
1E 2259+586	&	6.9790427250(15)	&	0.0483695(63)	&	0.59	&	17	&	-22.98$^{+0.47}_{-0.31}$	$^{**}$ &	944	&	3 $^{(d)}$ (Eph A,B2,D)	\\
 \hline
  \end{tabular}}
\begin{flushleft}
$\bf{Notes.}$ 

$^{(a)}$  $P, \dot{P}, B_{\mathrm{surf}}$ and $L_{\mathrm{x}}$ values are taken from \textit{McGill Online Magnetar Catalog} \footnotemark{} \citep{olausen2014}.

$^{(b)}$ Noise levels are measured in this work. Time span column gives the length of time for the noise measurement. 
$^{*}$ Noise measurements are previously presented by \cite{serim2012}.  
$^{**}$ Noise measurements are previously presented by \cite{cerri2016}. 

$^{(c)}$ Timing references are for frequency measurements taken from literature.

$^{(d)}$ Timing references are for timing solutions used in phase coherent timing analysis.

$^{(e)}$ Frequencies of 1E 1048.1--5937 are obtained from Fig. 5(a) of the given reference, then the lightcurves are folded at these frequencies.

$^{(f)}$ Frequencies of 1E 1547.0--5408 are obtained from Fig. 1(b) of the given reference, then the lightcurves are folded at these frequencies.

$^{(g)}$ The sampling rate of observations are not sufficient to measure the noise strength of these sources.

$\bf{References.}$ (1) \cite{mcgarry2005}; (2) \cite{tiengo2008}; (3) \cite{dib2014}; (4)  \cite{rea2013}; (5) \cite{camero2014}; (6) \cite{kulkarni2003}; (7) \cite{guver2012}; (8) \cite{kuiper2012}; (9) \cite{rodriguez2014}; (10) \cite{sato2010}; (11) \cite{halpern2010}; (12) \cite{coti2015}; (13) \cite{mere2000}; (14)\cite{woods2002}; (15) \cite{kaplan2002}; (16) \cite{mere2005};  (17) \cite{tiengo2005}; (18) \cite{woods2007}; (19) \cite{mere2007}; (20) \cite{esposito2007}; (21) \cite{nakagawa2009}; (22) \cite{younes2015}; (23) \cite{israel2003}; (24) \cite{gotthelf2004}; (25) \cite{ibrahim2004}; (26) \cite{halpern2005}; (27) \cite{gott2005}; (28) \cite{camilo2007}; (29) \cite{hotan2007}; (30) \cite{bernardini2009}; (31) \cite{camilo2016}; (32) \cite{scholz2014}; (33) \cite{esposito2011}; (34) \cite{esposito2013}; (35) \cite{rea2014}; (36) \cite{zhou2014}; (37) \cite{israel2016} 

\end{flushleft}
\end{table*}

If the pulse frequency history of a source is already presented in literature, we directly use the frequencies for noise strength measurements (see Table \ref{sources} and references therein).
If timing ephemerids are presented, we use the following approach to generate the pulse frequency histories.
Using the process described in Section 2, we extracted lightcurves for each source and the time series are corrected for solar system barycenter.
\footnotetext{http://www.physics.mcgill.ca/$\sim$pulsar/magnetar/main.html}
We employed phase coherent timing technique to determine the rotational phase as a function of time $\phi(t)$.
A pulse profile for each observation is generated by folding the light curves with pulse frequency of the source.
The template pulse is determined as the one with maximum $\chi^2$ among all pulse profiles.
Then, pulse profiles are represented in terms of harmonic functions \citep{deeter1982} and cross-correlated with the template.
As a result of cross-correlation, arrival times of pulses (pulse TOAs) are calculated for each observation. 
The rotational phase of a pulsar as function of time can be expressed as Taylor Expansion;
\begin{equation}
 \Phi(t) = \Phi_{0} + \nu (t-t_{0}) + \frac{1}{2} \dot{\nu} (t-t_{0})^2 + ....
\end{equation}
where $\nu$ is the pulse frequency at folding epoch $t_{0}$ and $\dot{\nu}$ is the pulse frequency derivative.
After constructing pulse time of arrivals, we splited it into many overlapping short time intervals.
Each interval contain different number of pulse TOAs depending on how frequently the source was observed. Then, we fit a linear model pulse TOAs within each time interval to measure pulse frequencies. 
The slope of these linear model, $\delta\phi = \delta\nu(t-t_0)$, allow us to estimate the pulse frequency corresponding to the mid-times of these intervals.
Uncertainties in the pulse frequency measurements are obtained using the error range of the slope of the linear model fit within the corresponding time interval 
and the horizontal error bars indicate the length of each time interval.
The measured pulse frequencies for each source is shown in Figure 1.

\begin{figure*}
  \centering{
  \hspace{0.5 cm}  \includegraphics[width=5.4cm, angle=270]{freqs/CXOUJ0100.eps} \hspace{0.4 cm} \includegraphics[width=5.4cm, angle=270]{freqs/4U0142.eps}
  
  \vspace{0.2 cm}
  
  \hspace{0.5 cm} \includegraphics[width=5.4cm, angle=270]{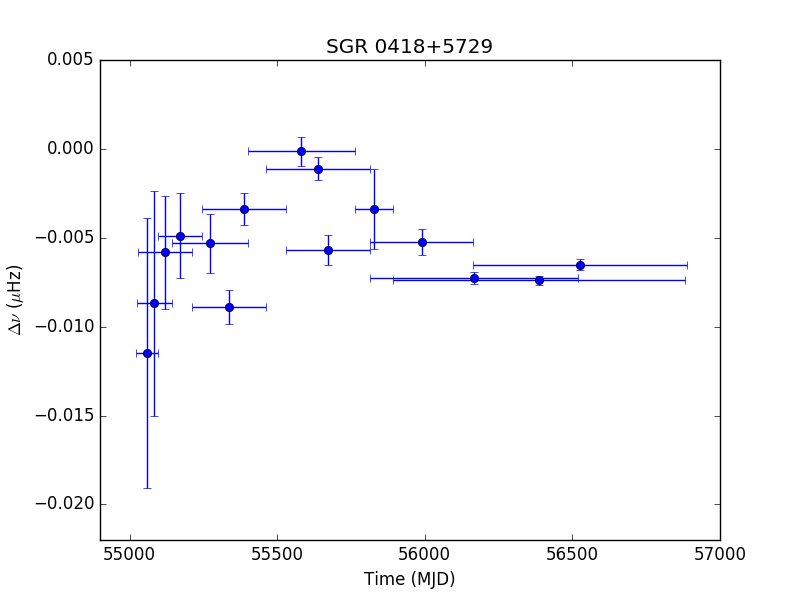} \hspace{0.5 cm} \includegraphics[width=5.4cm, angle=270]{freqs/SGR0501.eps}
  
  \vspace{0.2 cm}
  
  \includegraphics[width=5.4cm, angle=270]{freqs/SGR0526.eps} \hspace{0.7 cm} \includegraphics[width=5.4cm, angle=270]{freqs/CXOUJ1647.eps}
  
  \vspace{0.2 cm}
  
  \hspace{0.5 cm}  \includegraphics[width=5.4cm, angle=270]{freqs/1RXSJ1708.eps} \hspace{0.5 cm} \includegraphics[width=5.4cm, angle=270]{freqs/CXOUJ1714.eps}}
  \caption{Frequency histories of magnetars. Frequency measurements are taken from literature for CXOU J010043.1--721134, SGR 0526--66, CXOU J171405.7--381031,SGR 1806--20, XTE J1810--197, 3XMM J185246.6+003317 and SGR 1900+14 (see Table \ref{sources} for references). Frequencies of other sources are measured in this work. For two of the sources (SGR 0418+5729 and Swift J1822.3-1606) $\Delta\nu$ are plotted instead of $\nu$ since their spin-down rates are low.}
  \label{freqs}
\end{figure*}

 \begin{figure*}
 \ContinuedFloat
  \centering{  
  \includegraphics[width=5.4cm, angle=270]{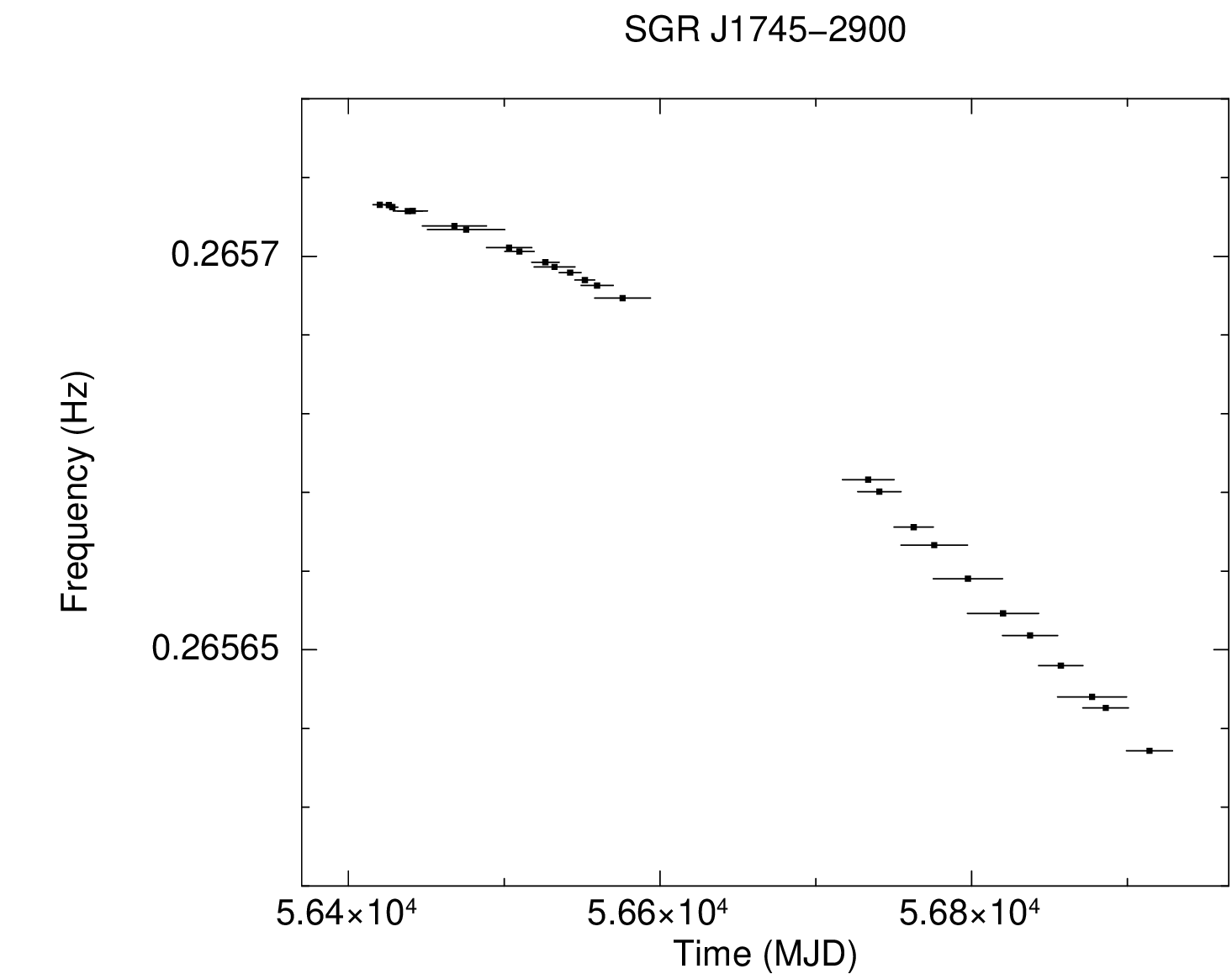} \hspace{0.5 cm} \includegraphics[width=5.4cm, angle=270]{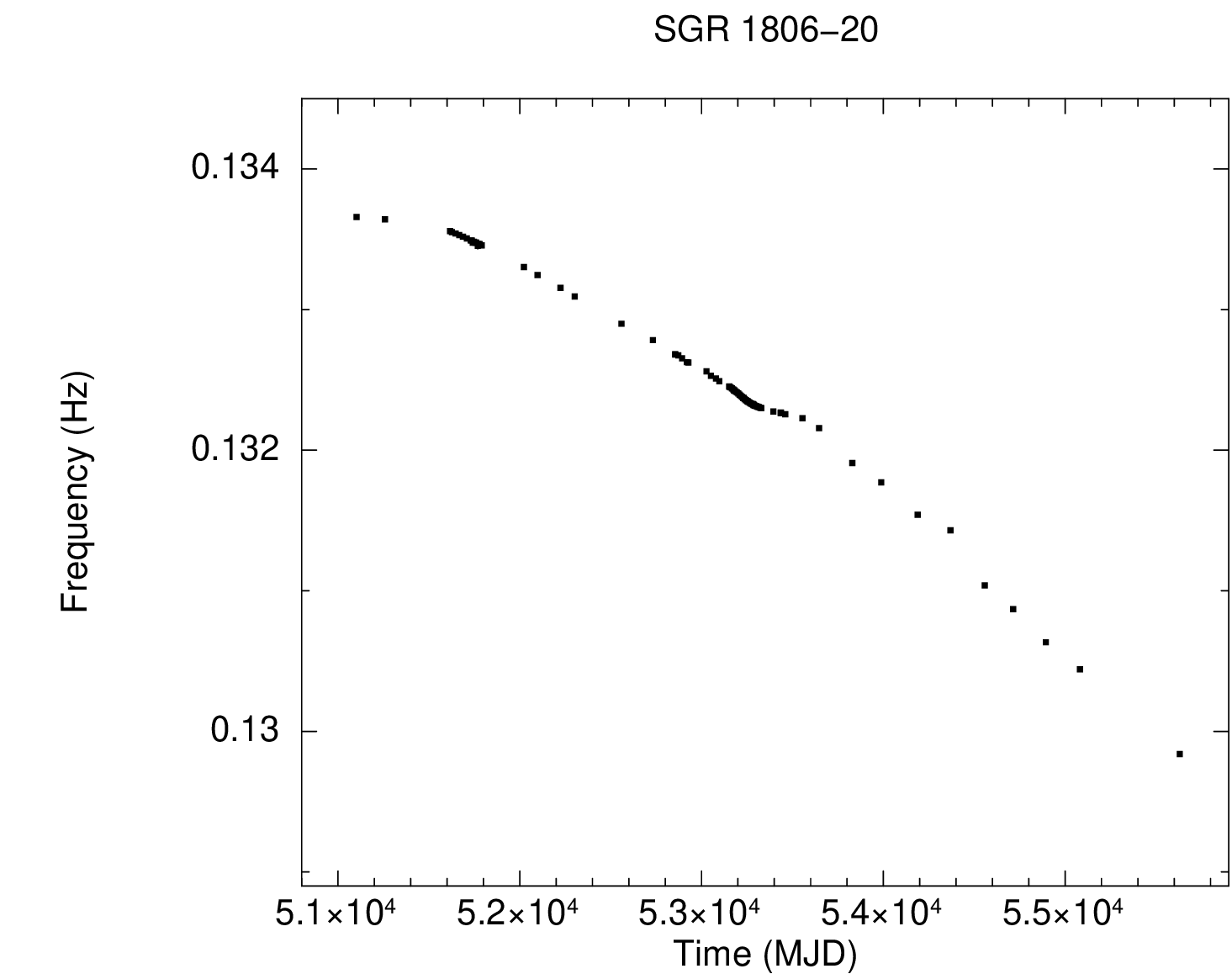}
  
  \vspace{0.2 cm}
  
  \hspace{0.3 cm} \includegraphics[width=5.4cm, angle=270]{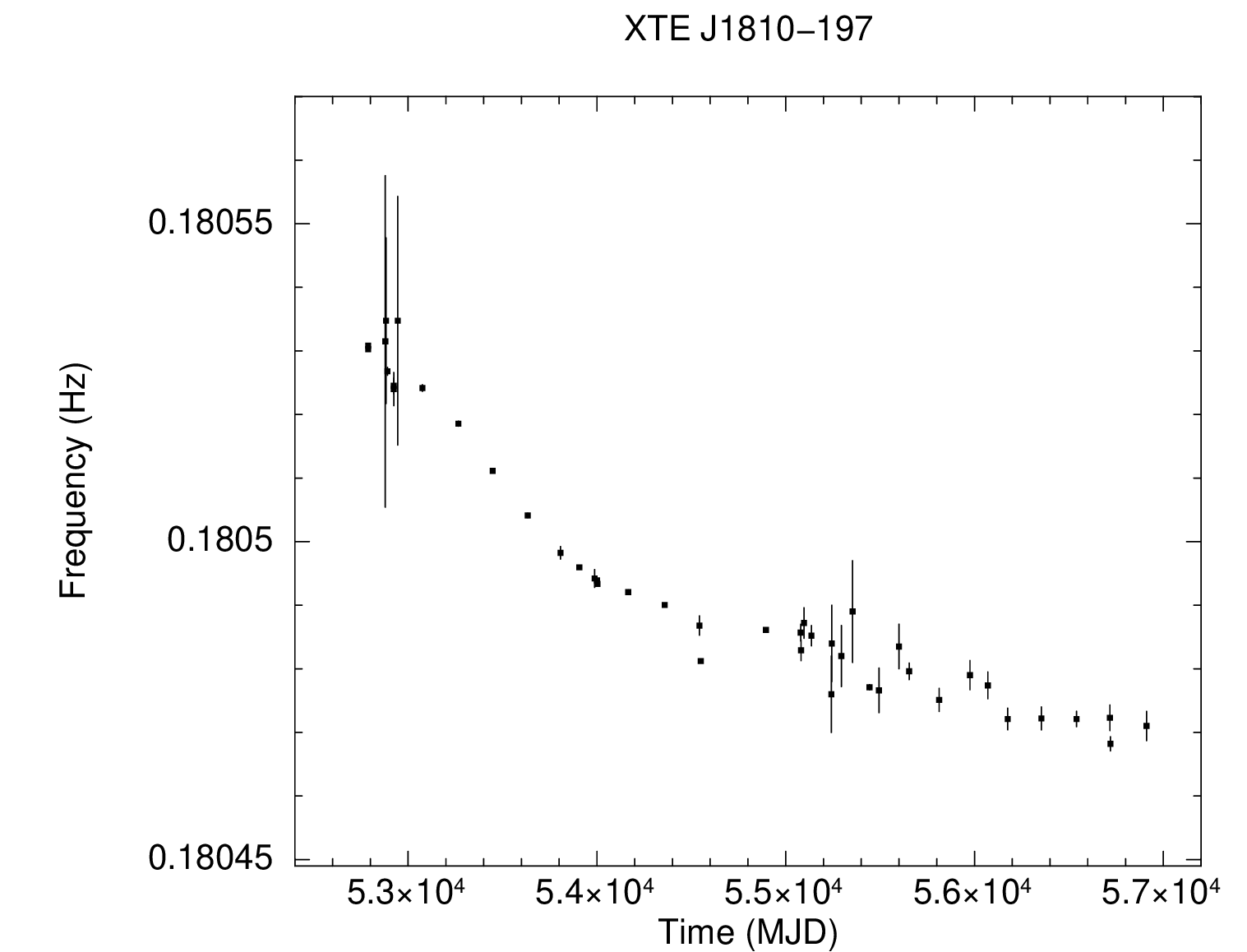} \hspace{0.7 cm} \includegraphics[width=5.4cm, angle=270]{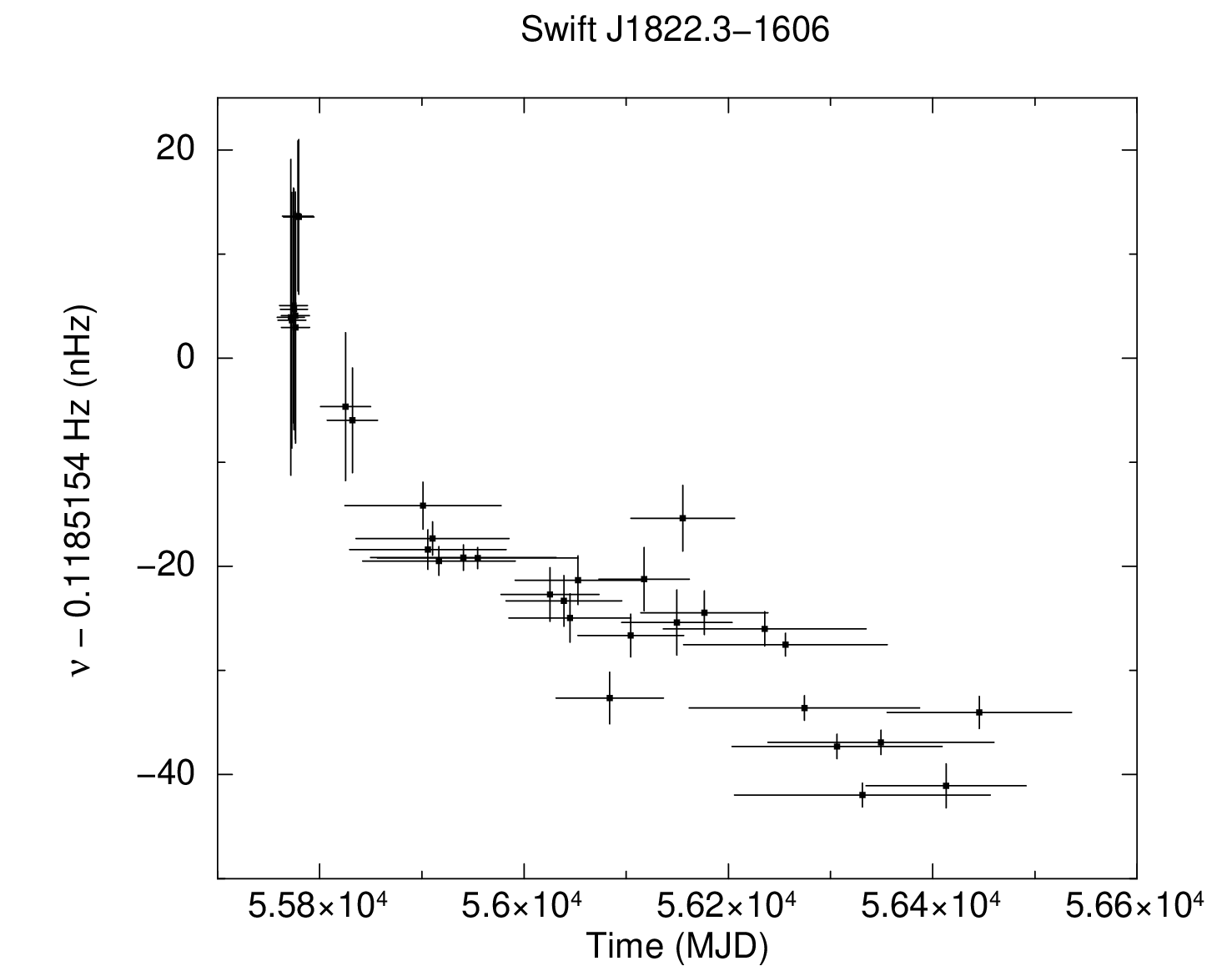}
  
  \vspace{0.2 cm}
  
  \includegraphics[width=5.4cm, angle=270]{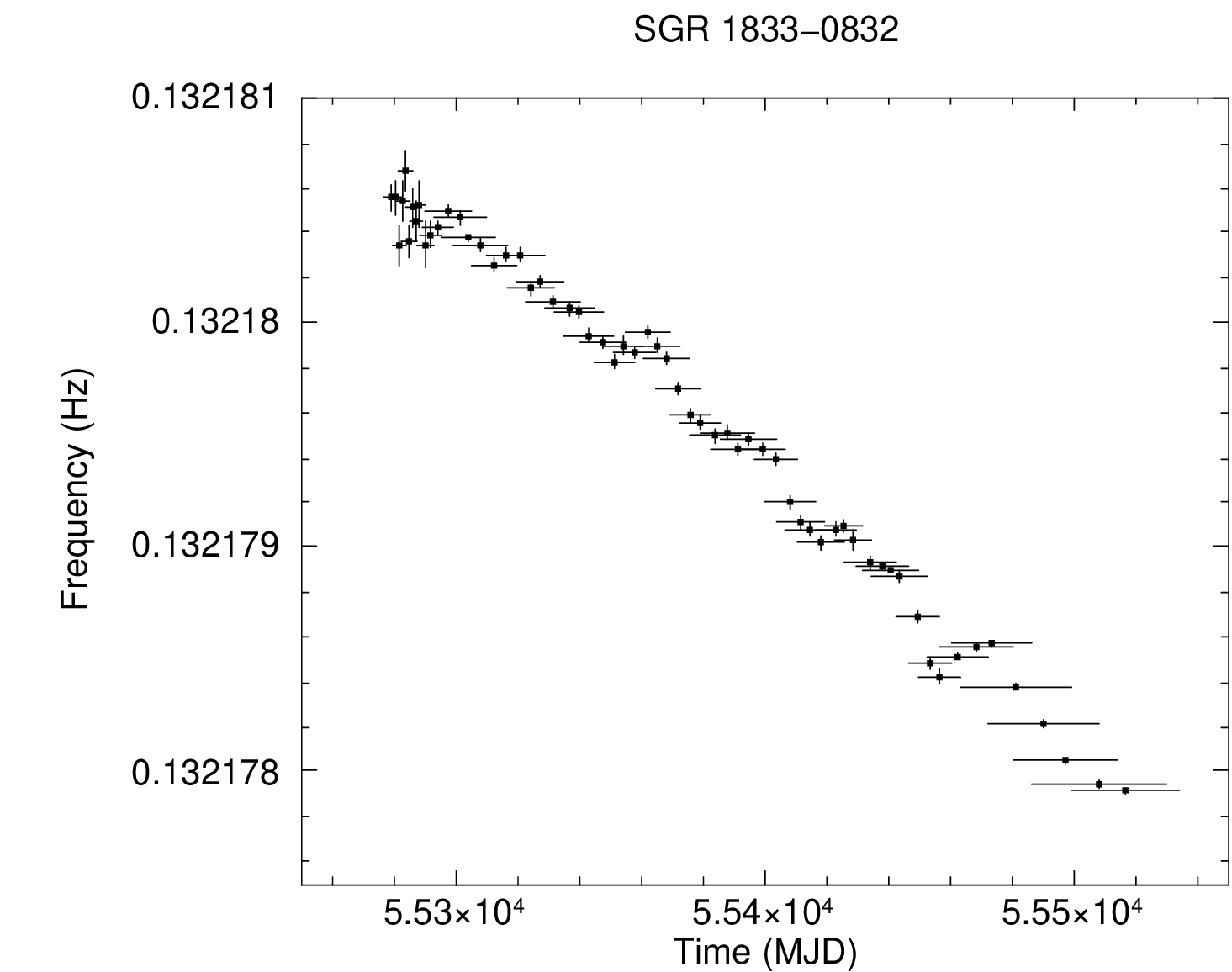} \hspace{0.5 cm} \includegraphics[width=5.4cm, angle=270]{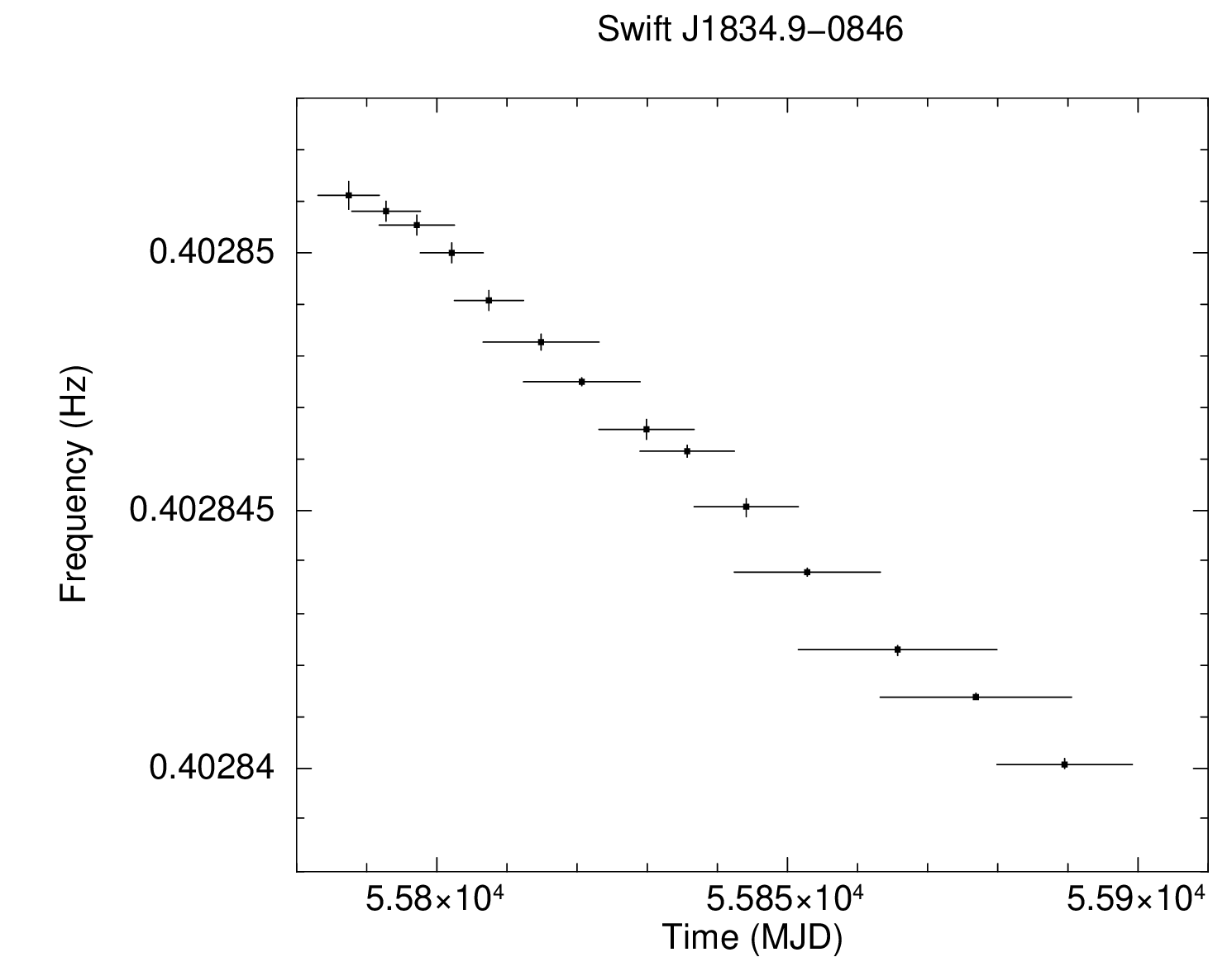}
  
  \vspace{0.2 cm}
  
  \hspace{0.2 cm} \includegraphics[width=5.4cm, angle=270]{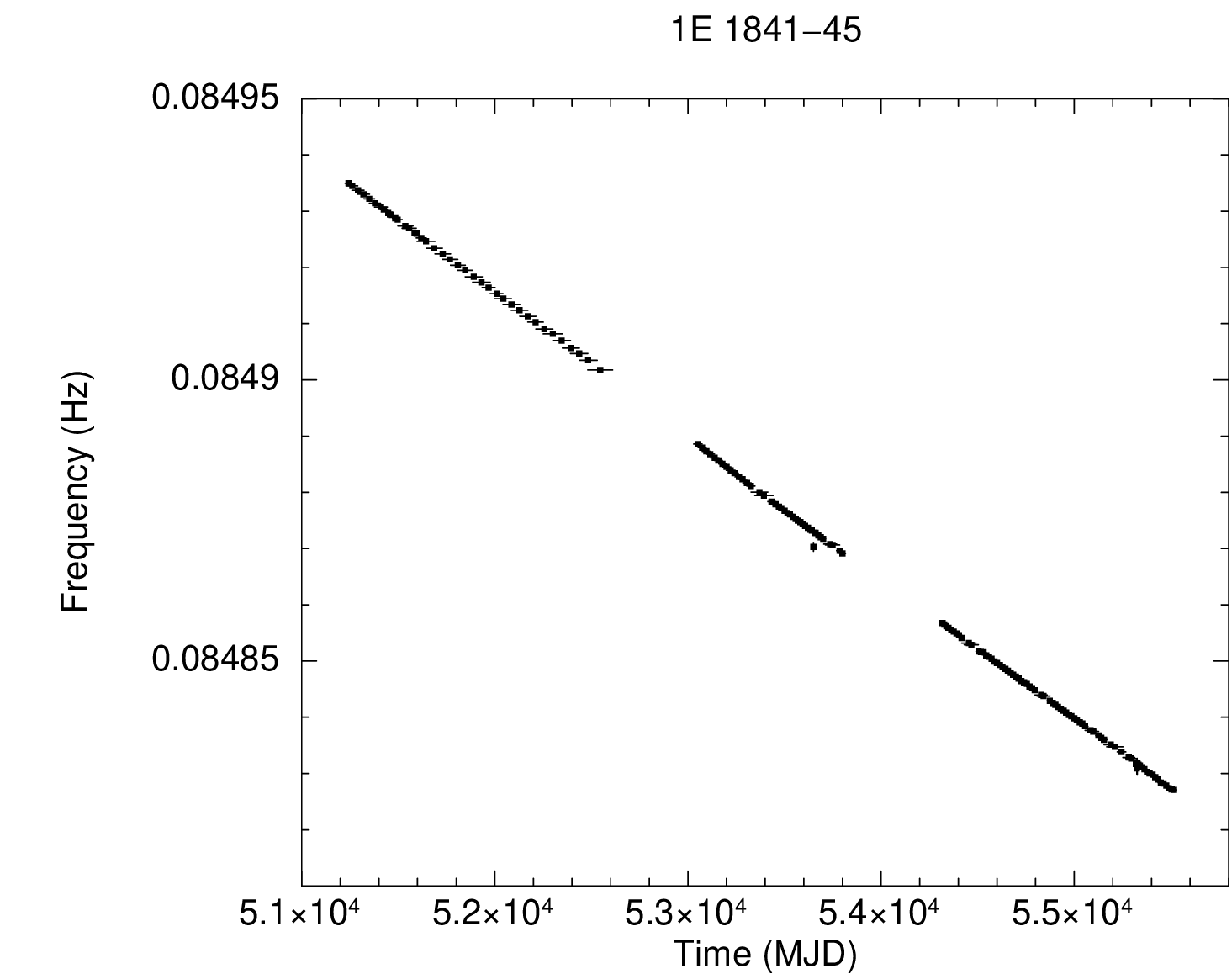} \hspace{0.5 cm} \includegraphics[width=5.4cm, angle=270]{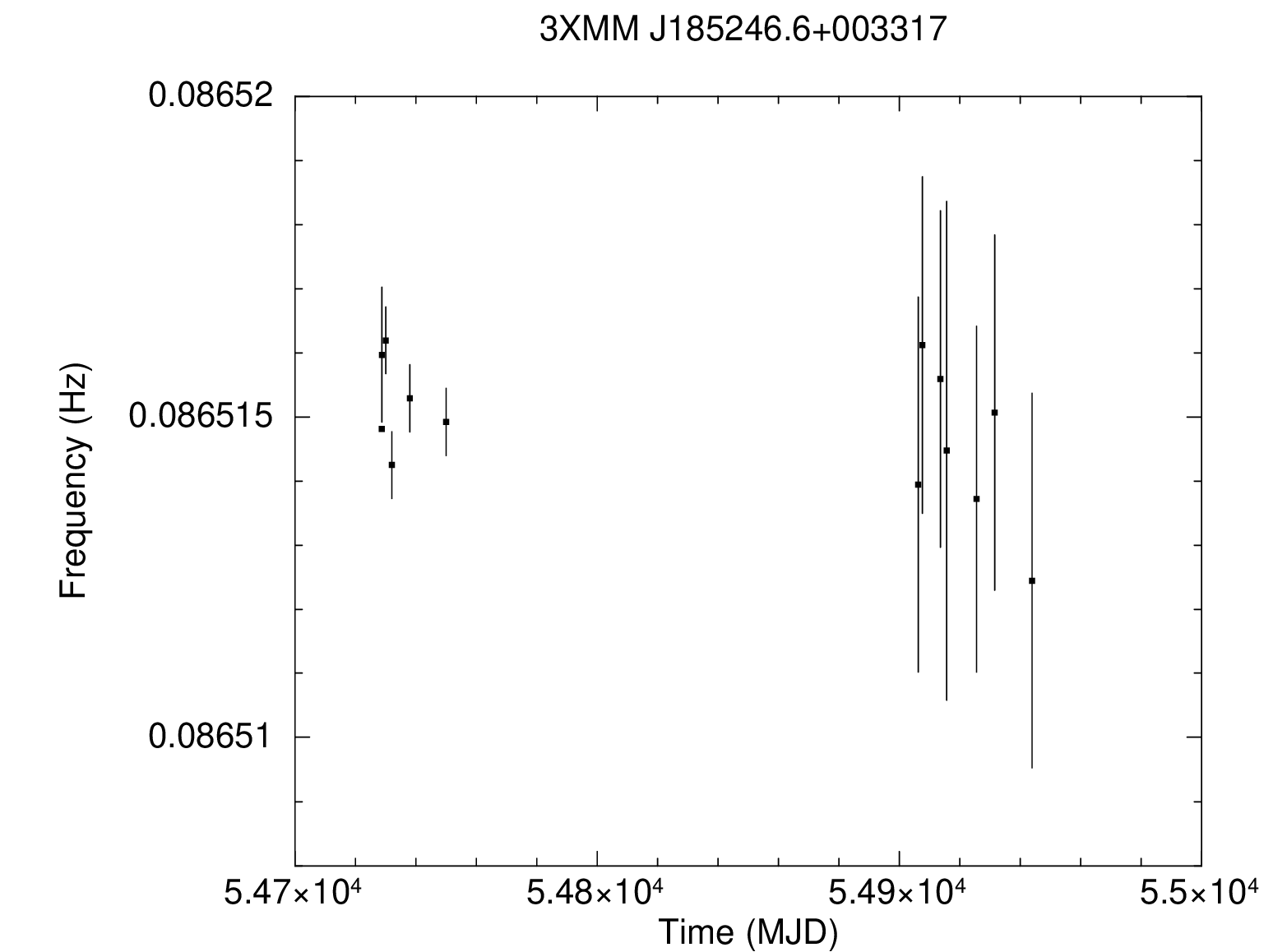}} 
  \caption{continued.}
  \label{freqs}
\end{figure*}

 \begin{figure*}
 \ContinuedFloat
  \centering{  
  \includegraphics[width=5.4cm, angle=270]{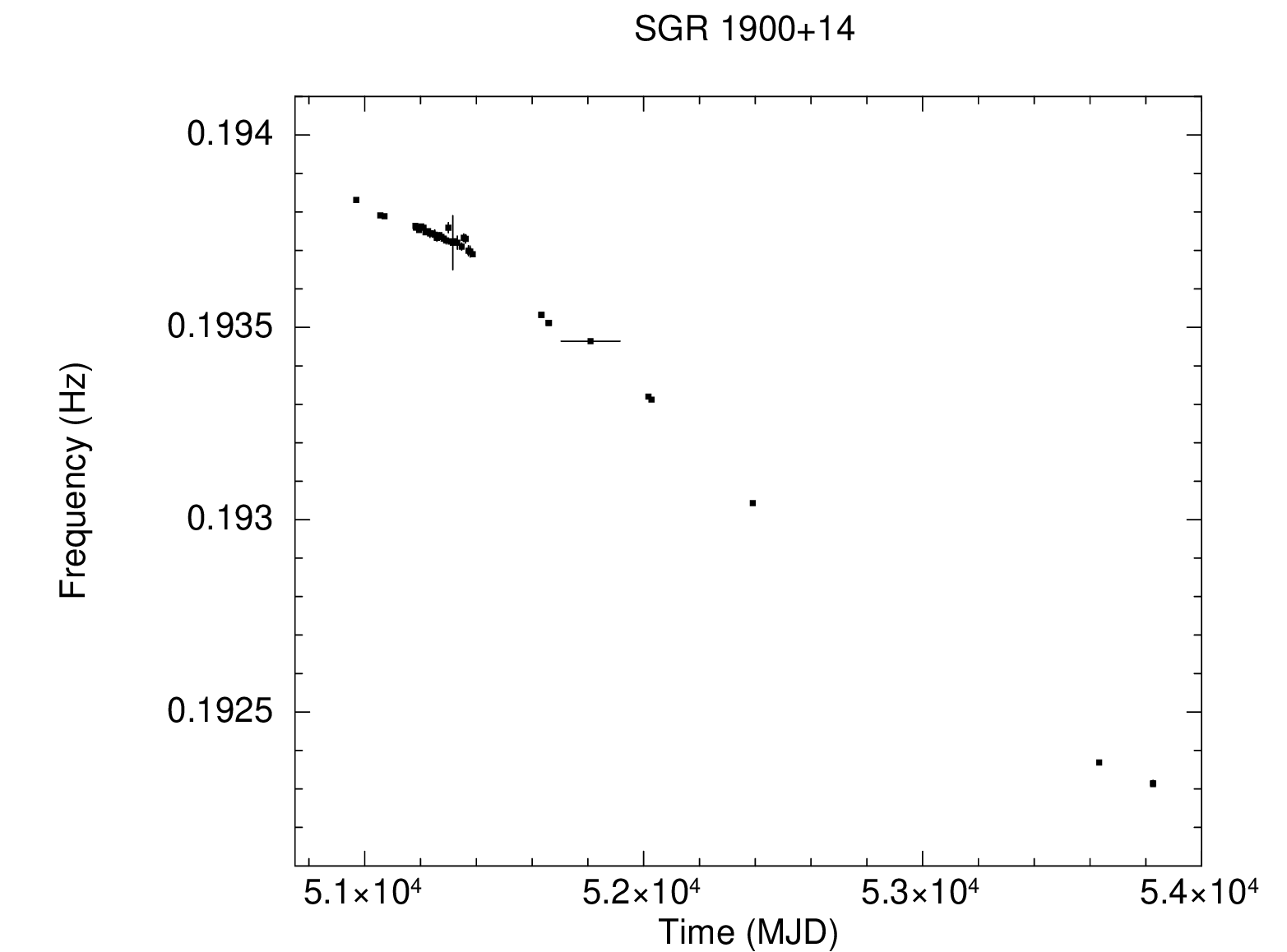} \hspace{0.5 cm}\includegraphics[width=5.4cm, angle=270]{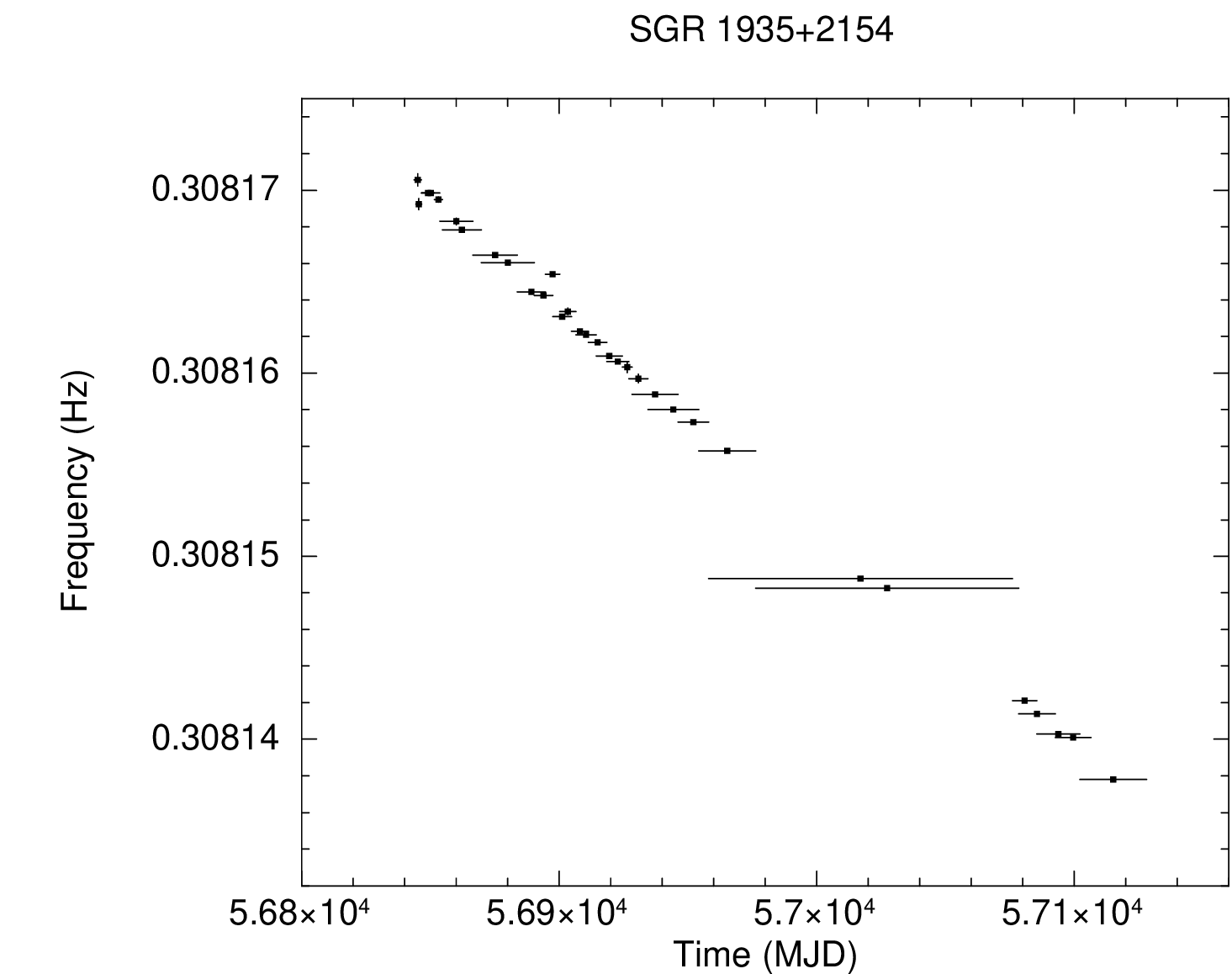} 
  
  \vspace{0.2 cm}
  
  \includegraphics[width=5.4cm, angle=270]{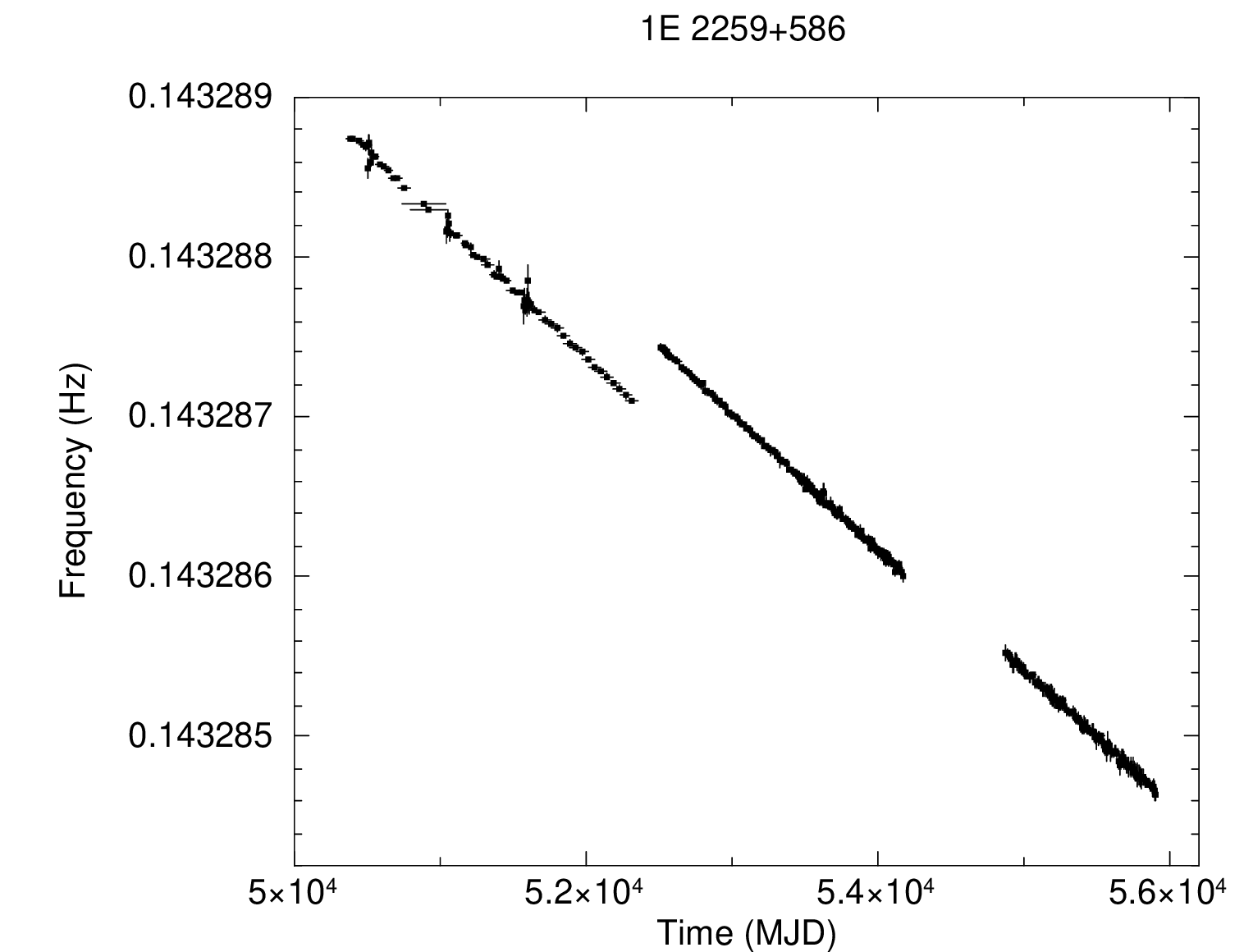}}
  \caption{continued.}
  \label{freqs}
\end{figure*}

In order to investigate torque noise characteristics, we investigated the rms of the residuals of pulse frequency histories for long time scales and pulse TOAs for shorter time scales (if a timing solution is available).
The calculated mean square residual value $<\sigma^2_R(m,T)>$, depends on the degree $m$ of polynomial removal and time span $T$ of data set. 
In our analysis, we remove simple spin-down trend (quadratic model (m=2) for pulse TOAs or linear model (m=1) for pulse frequency histories) for all magnetars.
Then, the power density spectra of pulse frequency derivative fluctuations are established by employing the root-mean-square (rms) residual technique \citep[see][for details]{cordes1980,deeter1984}.
In this technique, r$^{\mathrm{th}}$ order red noise strength $S_r$ over a time span $T$ can be estimated via;
\begin{equation}
 <\sigma^2_R(m,T)> = S_r T^{2r-1}  <\sigma^2_R(m,1)>_u
\end{equation}
where $<\sigma^2_R(m,1)>_u$ is the normalization factor for the unit noise strength.
The unit noise strength $<\sigma^2_R(m,1)>_u$ also depends on the degree of the polynomial removed before the calculation of rms value.
This normalization factor for the unit noise strength can be determined by finding the expected mean square residual for
unit strength red noise (S$_r$= 1) over a unit time interval (T = 1).
The normalization factor for the unit noise strengths can be obtained either by numerical simulation \citep{scott2003} or via direct calculation \citep{deeter1984}.
In our study, we obtained rth order noise strength $S_r$  using the normalization factors calculated by \citep[see Tables 1 and 2]{deeter1984}.
If the noise strength estimations are constant, independent of sampling frequency, the power spectra should have zero slope.
In this case, spin frequency fluctuations can be explained with the random walk model therefore fluctuations of spin frequency derivatives can be expressed as white noise (or delta function fluctuation at time).
Our preliminary analysis showed that the residuals of the spin frequencies can be
characterized by first order red noise (or random walk).
The order of red noise (r=1,2,3..) corresponds to the rth-time integral of white noise time series.
Therefore, in the construction of power density spectra, we use $r=1$ for spin frequency histories and $r=2$ for the pulse TOAs, respectively.
As a next step, the maximum time span of the available data set is considered as the longest time scale (T) and noise strength calculations are repeated for smaller time scales (T/2, T/4, ...).
The calculated noise strengths on each time scale are logarithmically averaged and combined into a single power estimate.
Finally, the power density spectra of pulse frequency derivative fluctuations are constructed using the power estimates on different time scales (with the corresponding analysis frequency  $f=1/T$).

\section{Results and Discussion}

In this study, noise characteristics of magnetars are investigated via power density spectra of pulse frequency derivatives. Power density spectra estimates of fifteen magnetars are given in Figure \ref{power}.

\begin{figure*}
  \centering{
  \hspace{0.1 cm} \includegraphics[width=5.4cm, angle=270]{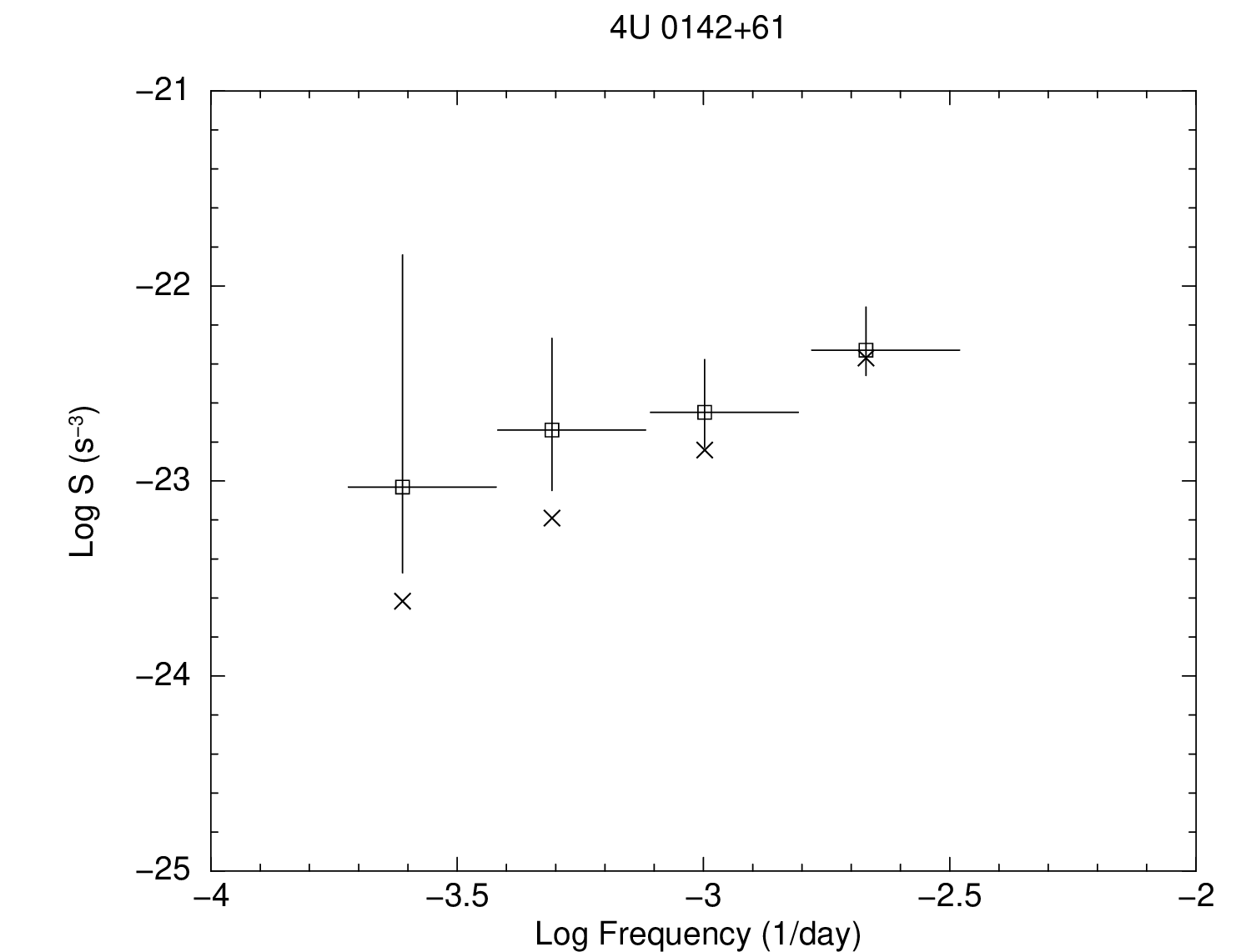} \hspace{0.5 cm}\includegraphics[width=5.4cm, angle=270]{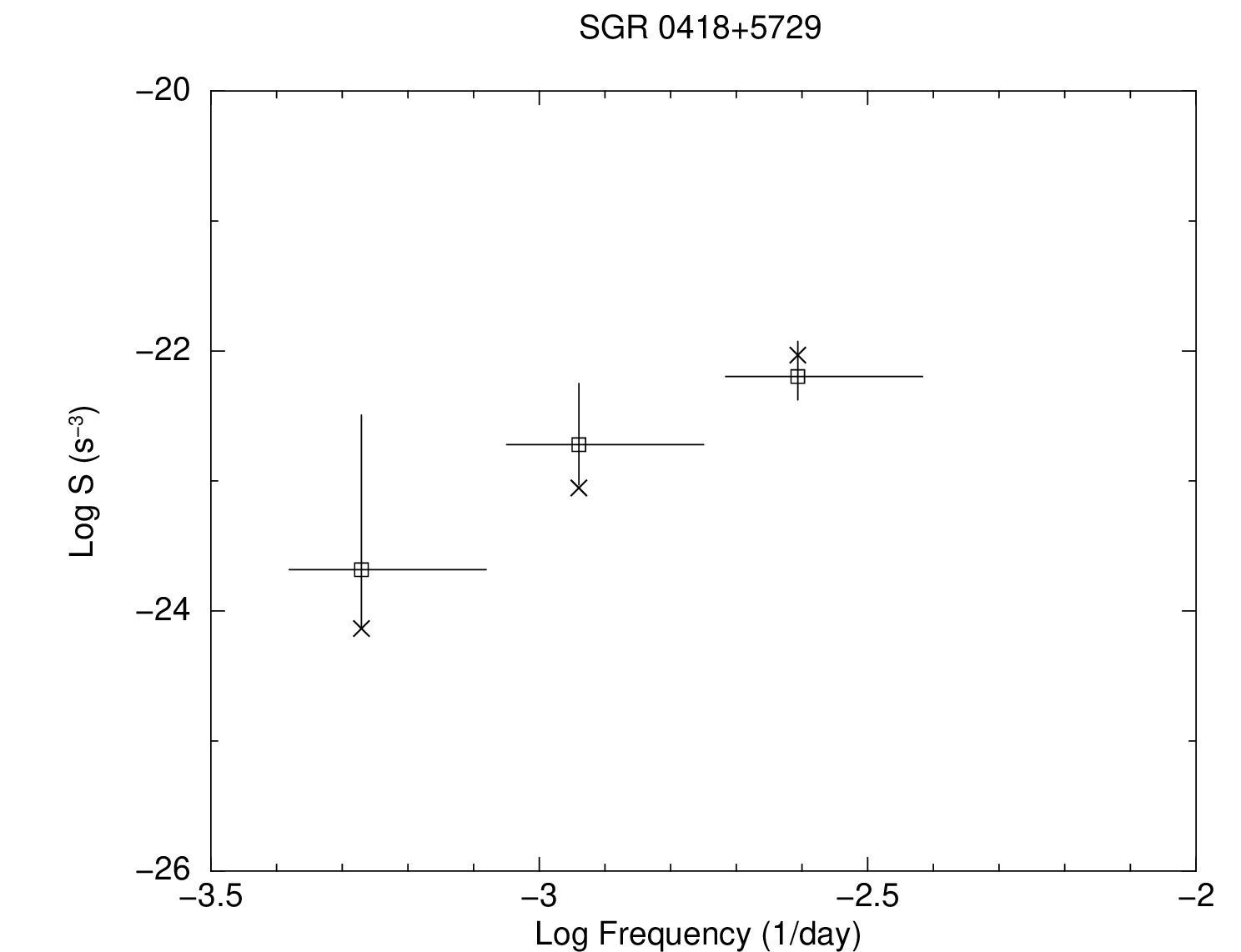} 
  
  \vspace{0.2 cm}
  
   \hspace{0.1 cm} \includegraphics[width=5.4cm, angle=270]{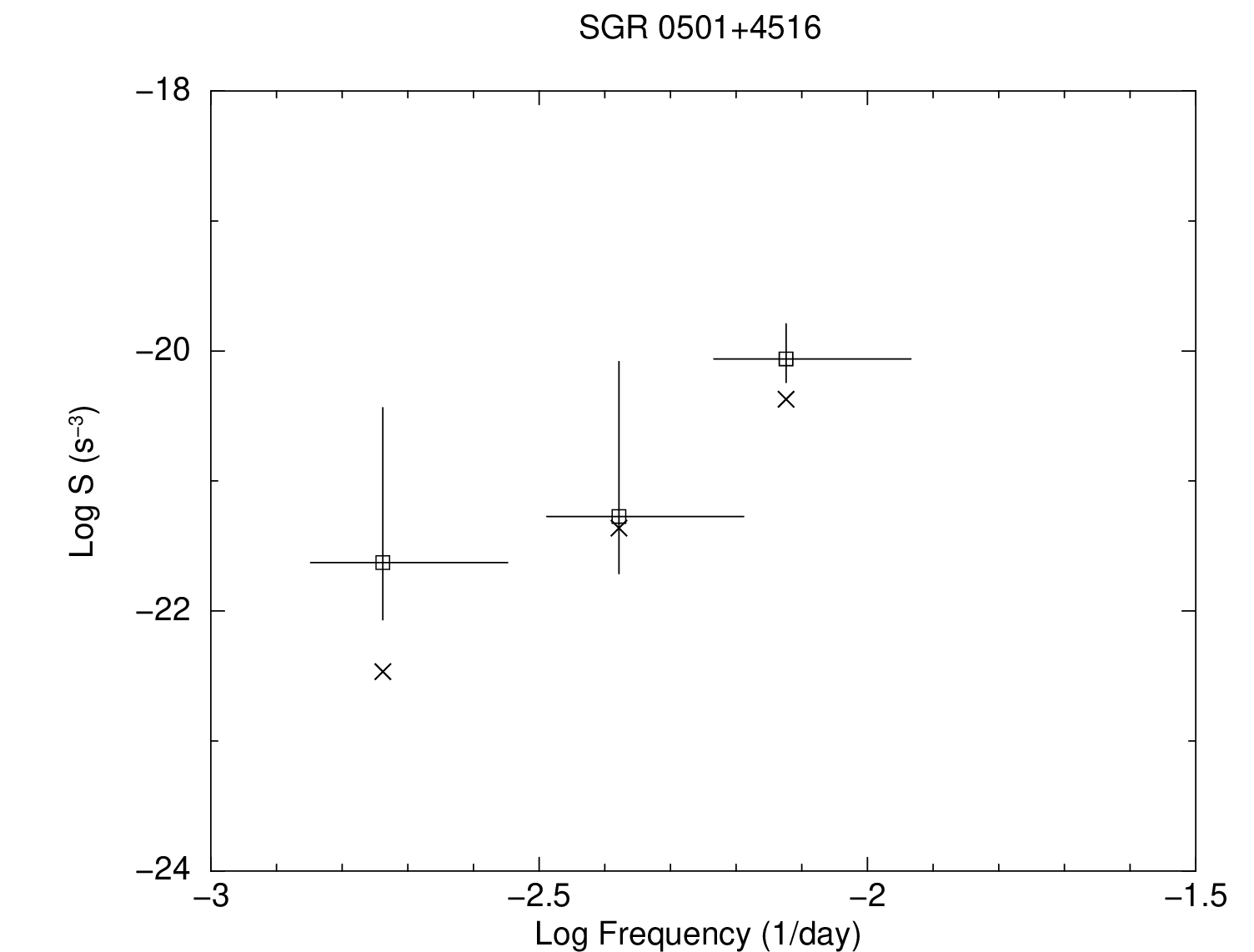} \hspace{0.4 cm}\includegraphics[width=5.4cm, angle=270]{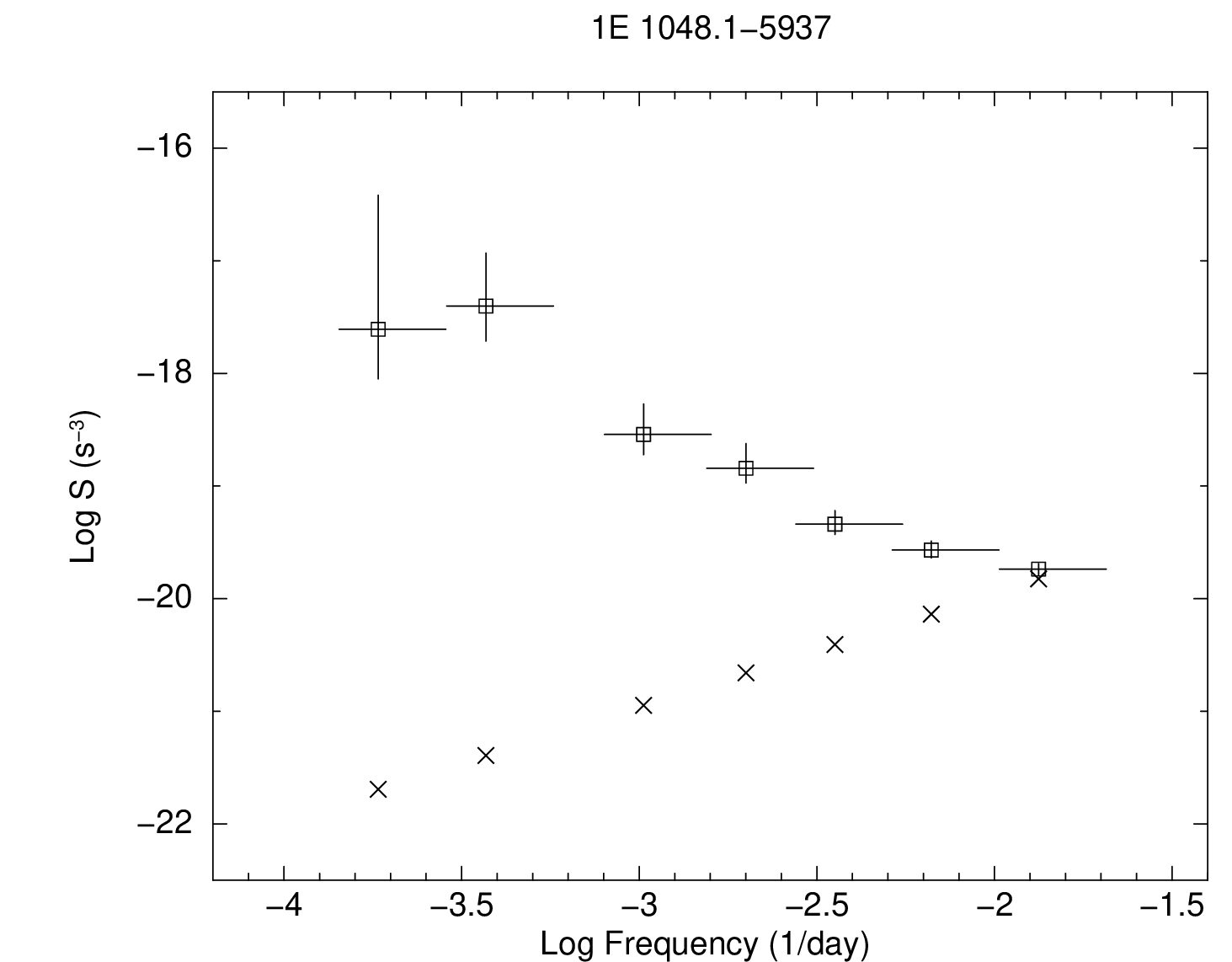} 
  
  \vspace{0.2 cm}
  
 \hspace{0.1 cm} \includegraphics[width=5.4cm, angle=270]{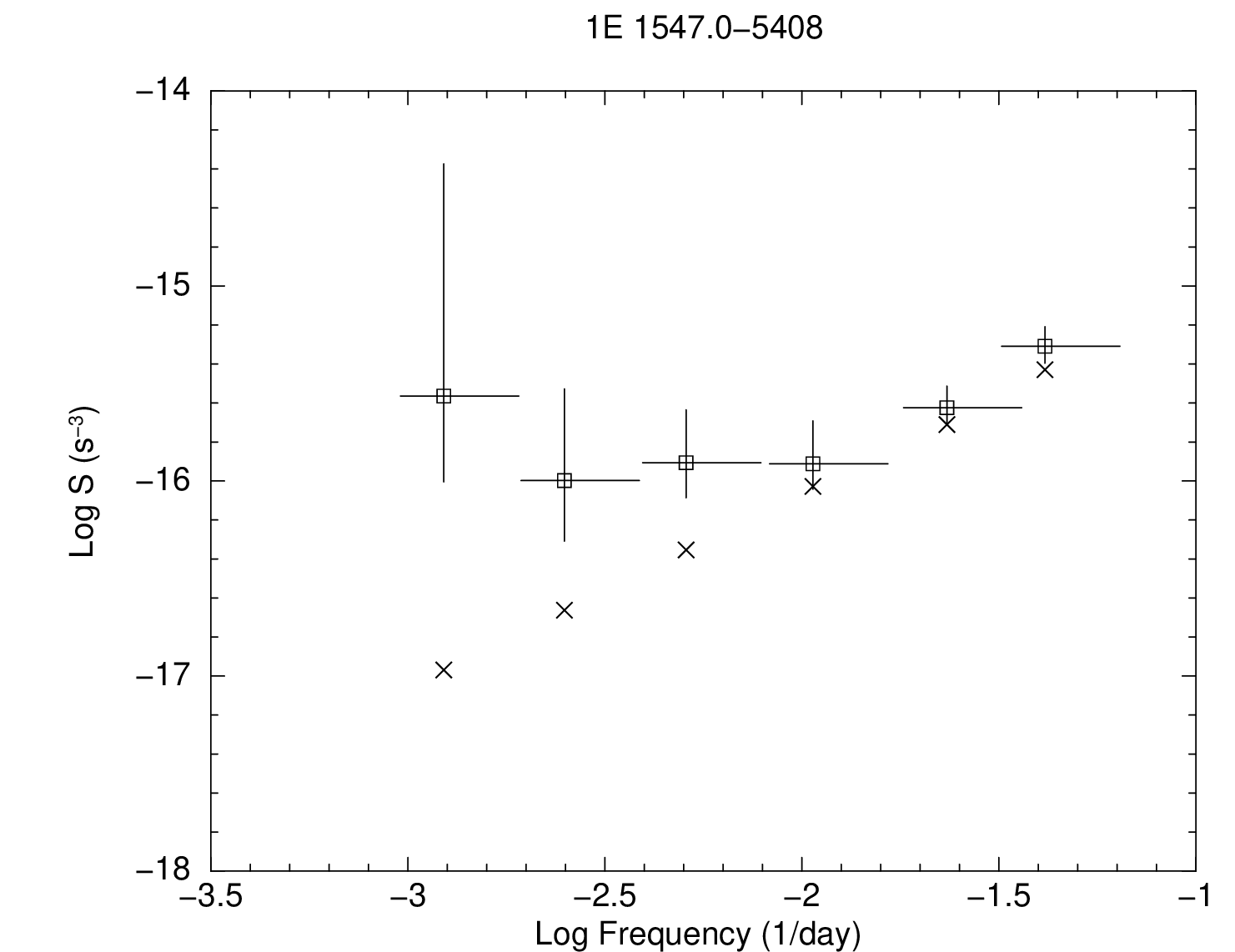} \hspace{0.5 cm}\includegraphics[width=5.4cm, angle=270]{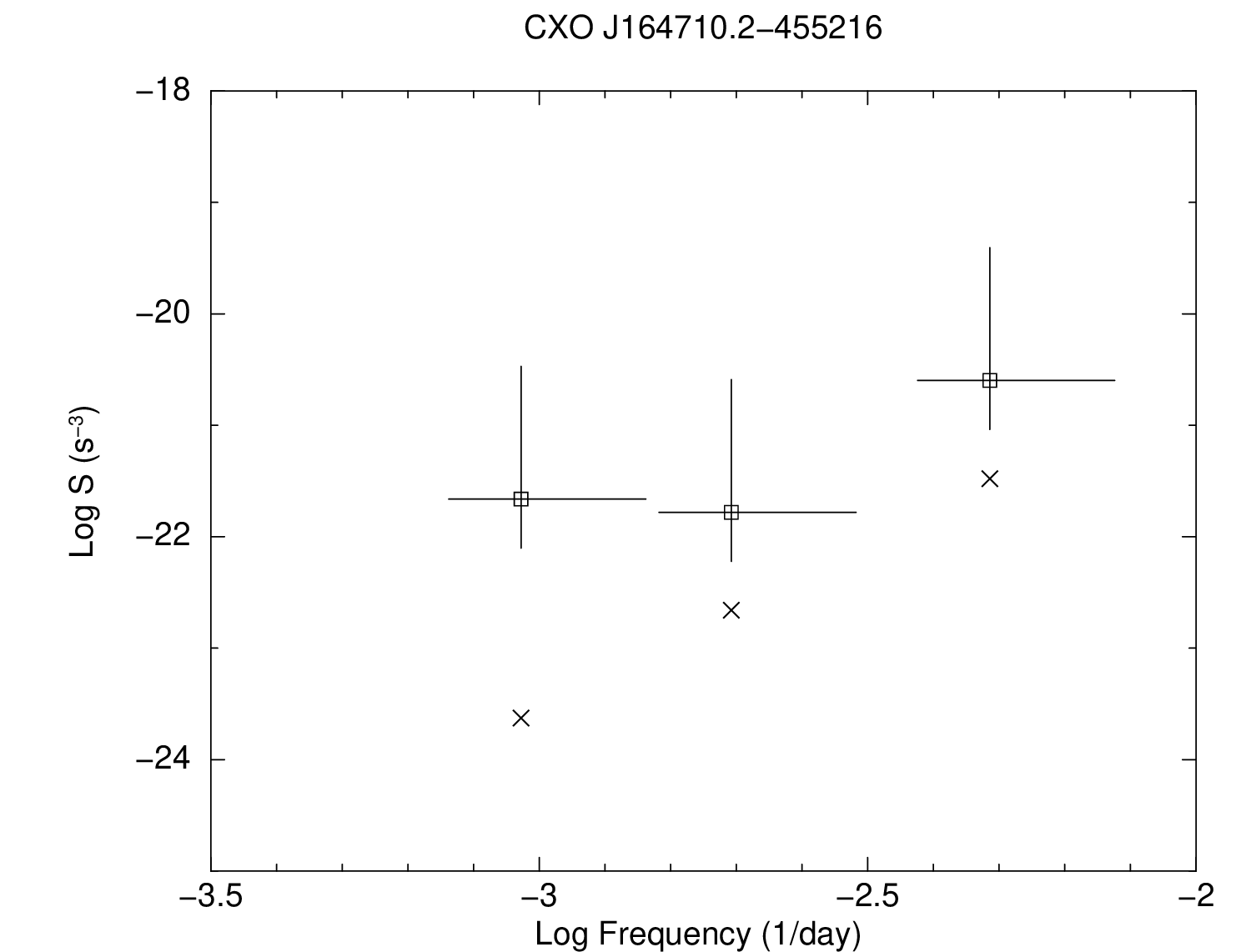}
  
  \vspace{0.2 cm}
  
  \includegraphics[width=5.4cm, angle=270]{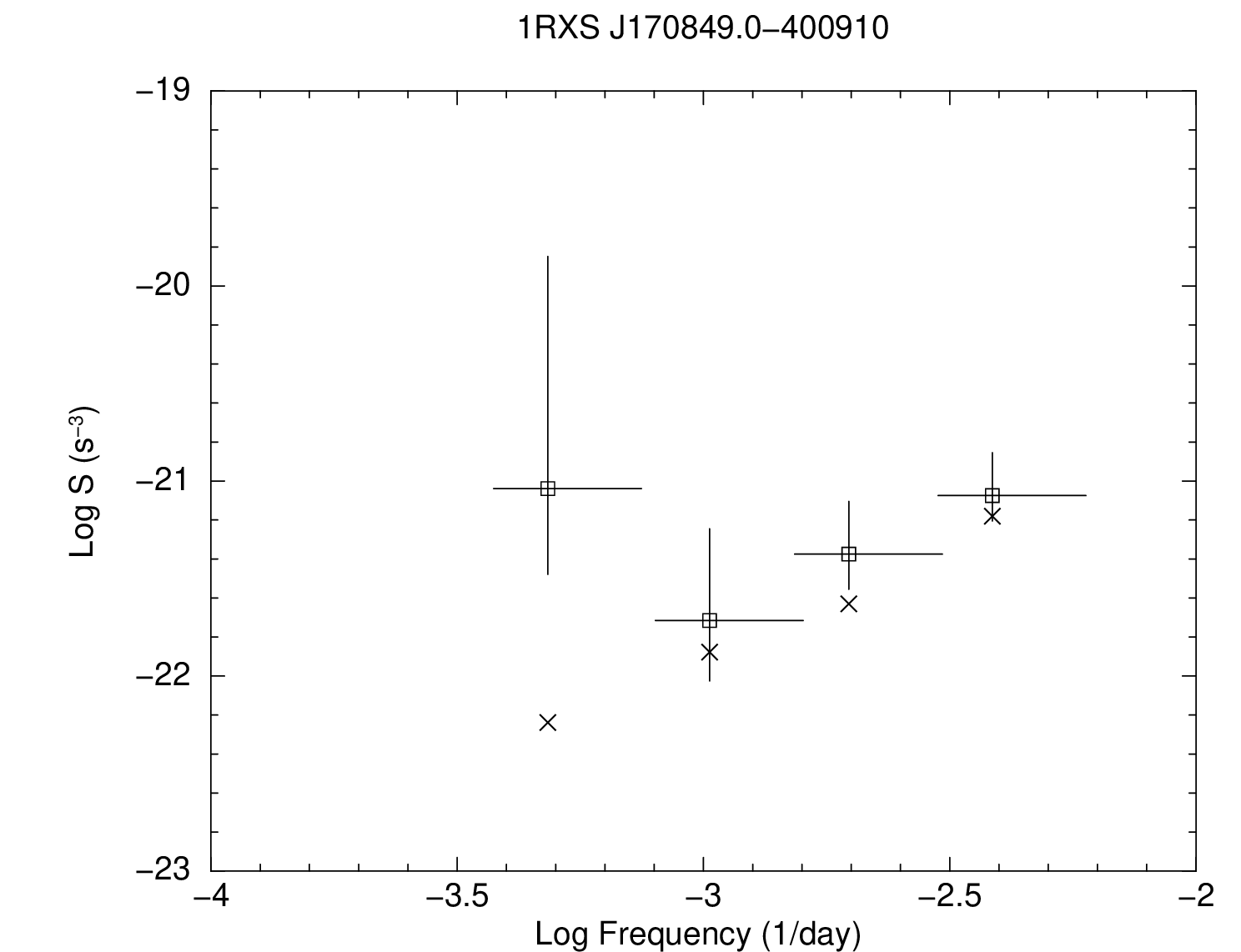} \hspace{0.5 cm}\includegraphics[width=5.4cm, angle=270]{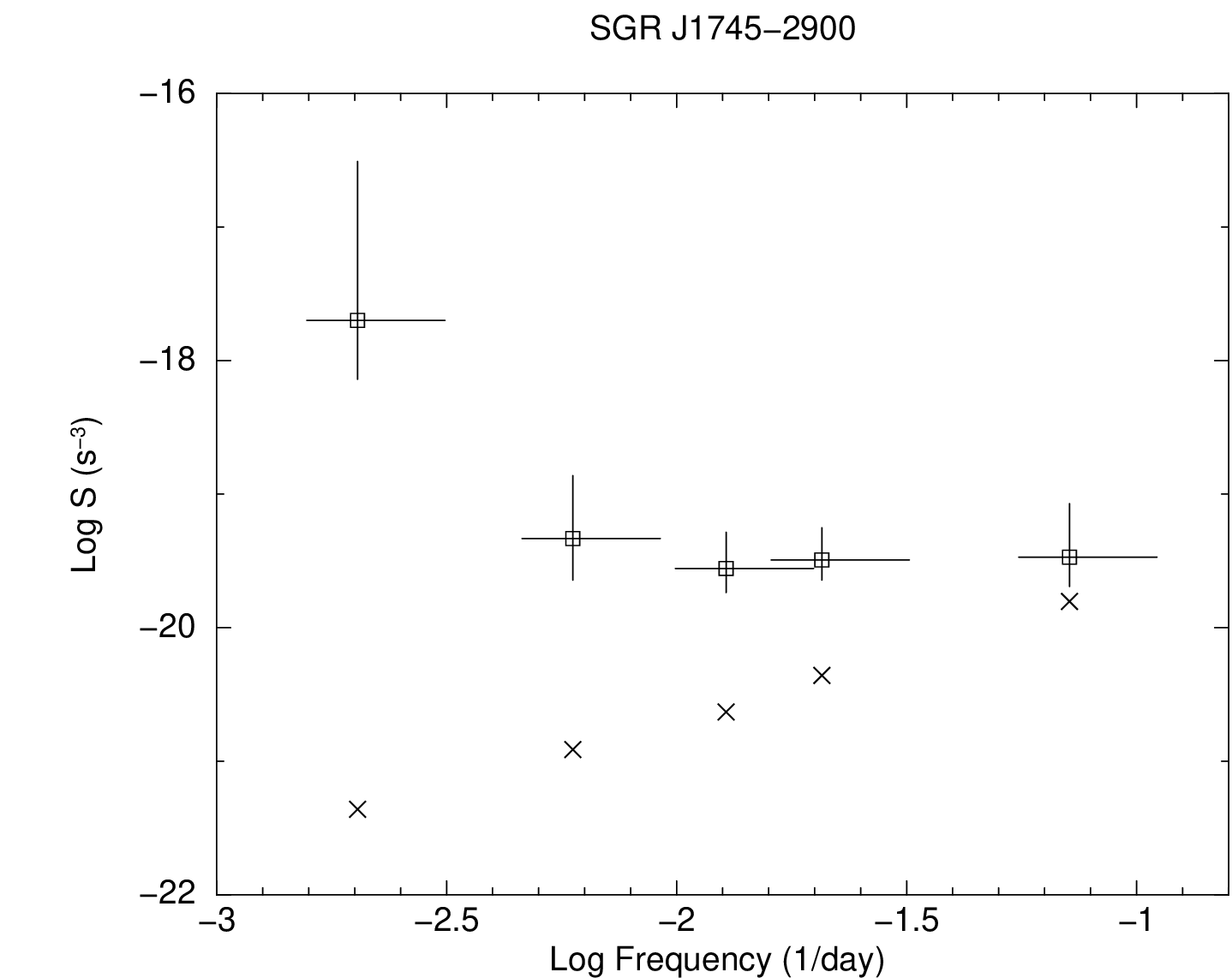}}
  \caption{Power density spectra of magnetars. Crosses indicate the power resulting from measurement noise.}
  \label{power}
\end{figure*}

\begin{figure*}
 \ContinuedFloat
  \centering{  
  \includegraphics[width=5.4cm, angle=270]{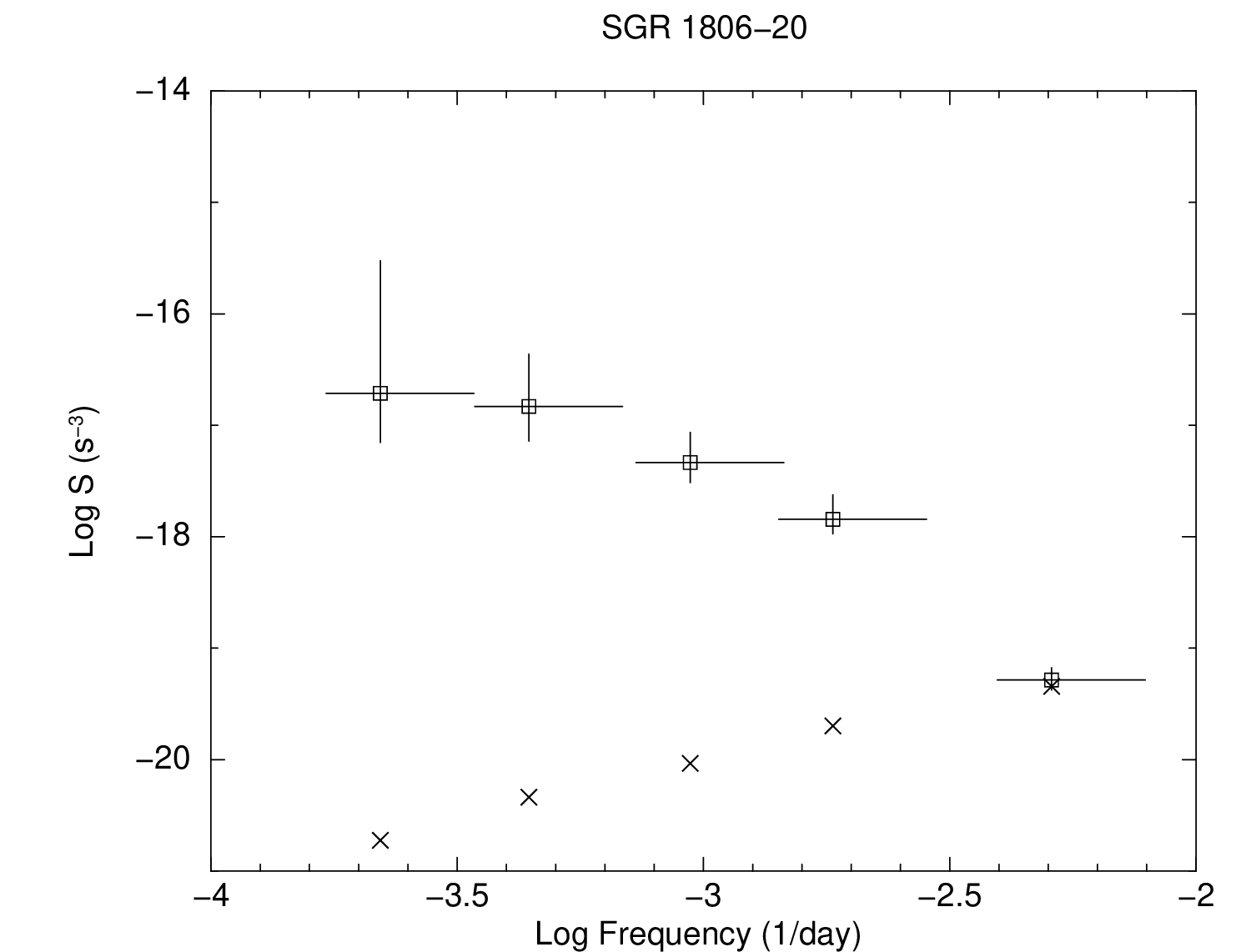} \hspace{0.5 cm}\includegraphics[width=5.4cm, angle=270]{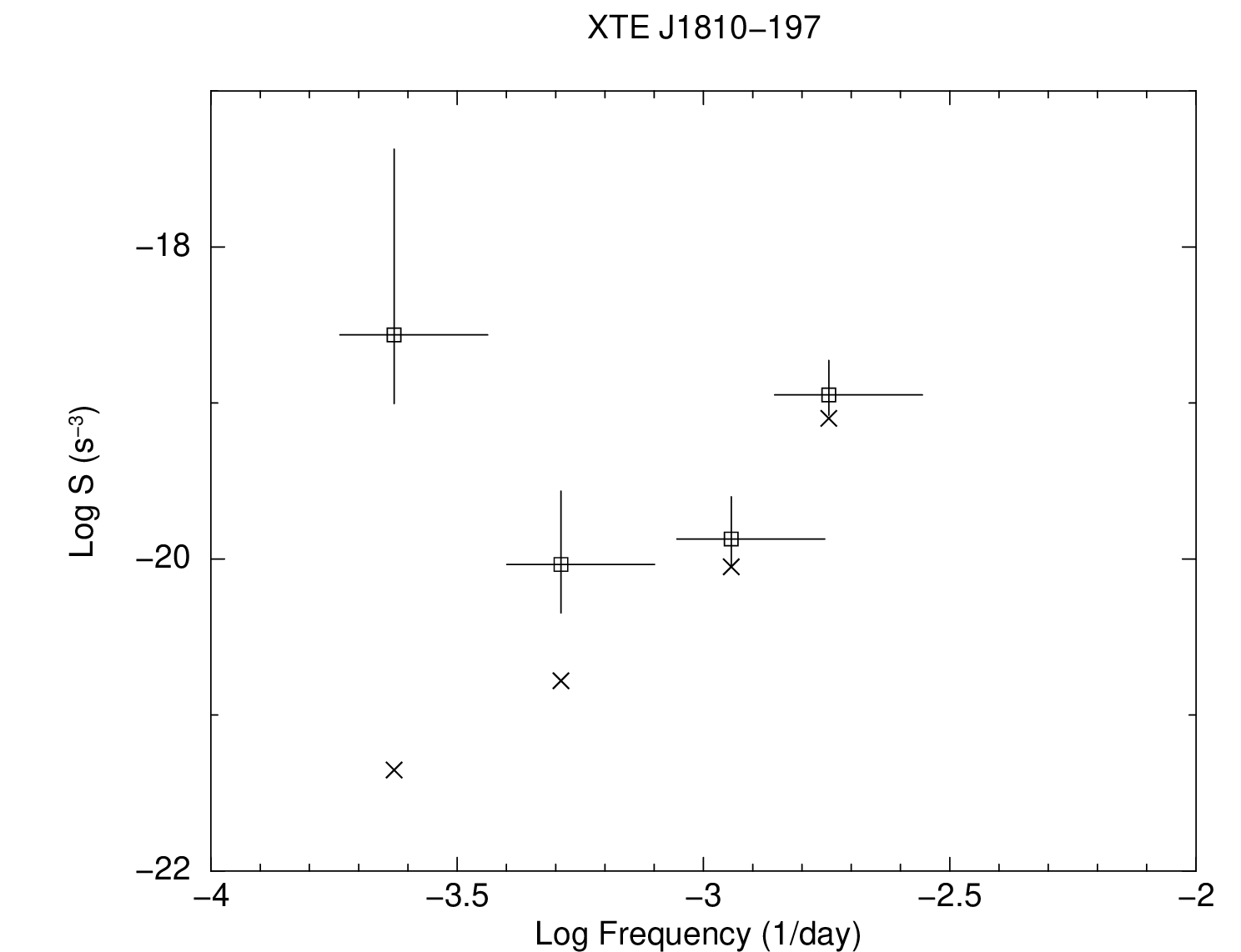} 
  
  \vspace{0.2 cm}
  
  \includegraphics[width=5.4cm, angle=270]{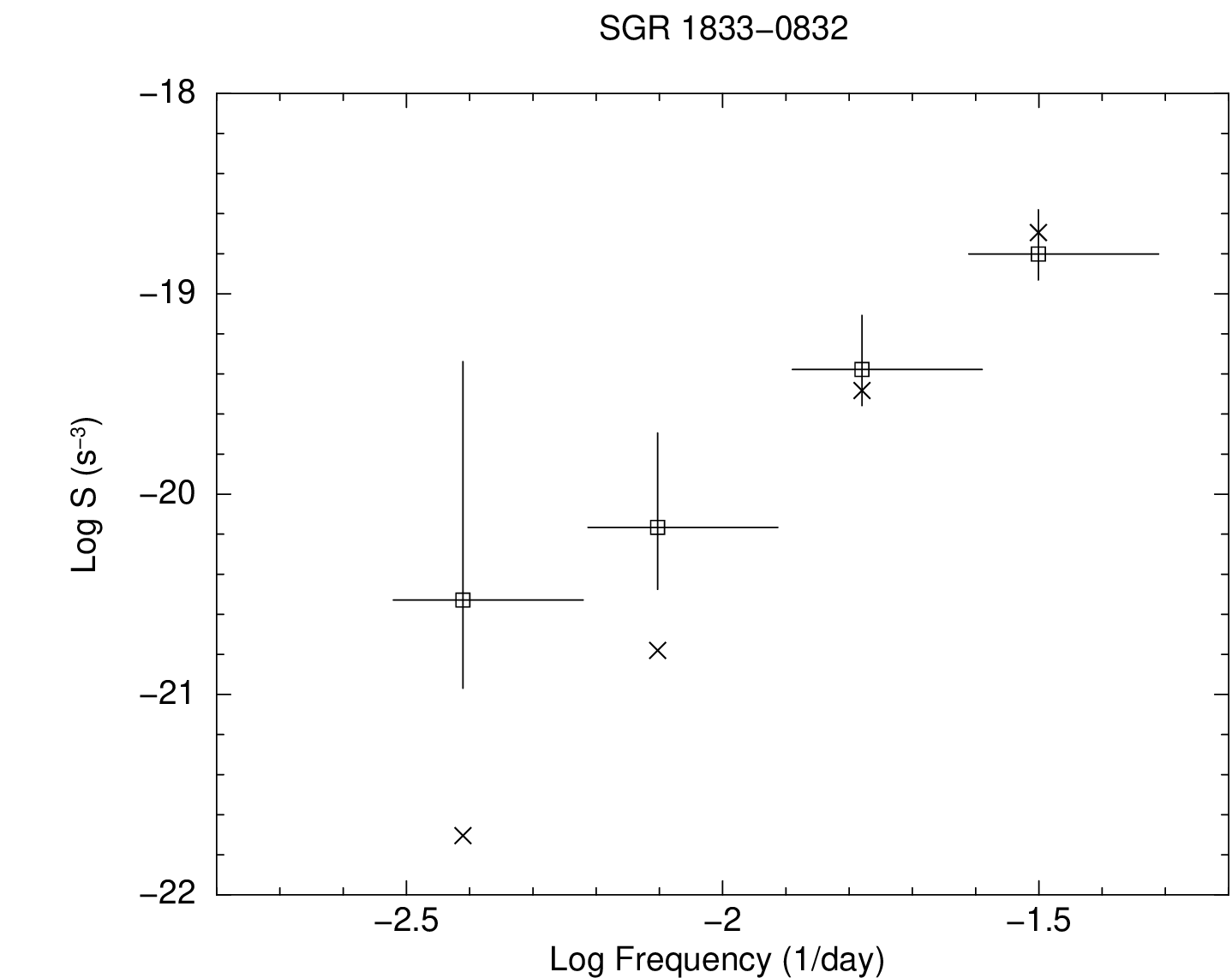} \hspace{0.7 cm}\includegraphics[width=5.4cm, angle=270]{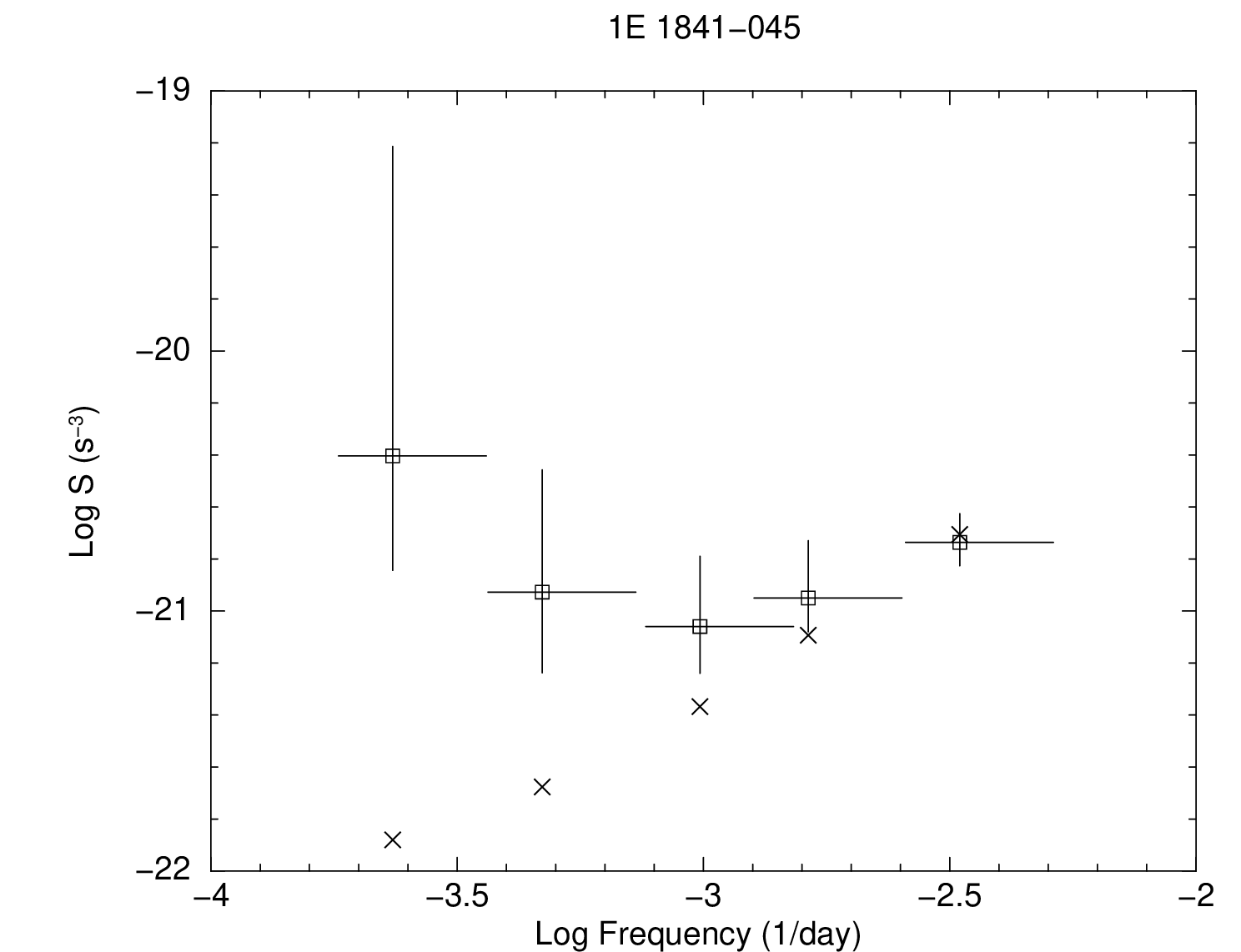} 
  
  \vspace{0.2 cm}
  
  \includegraphics[width=5.4cm, angle=270]{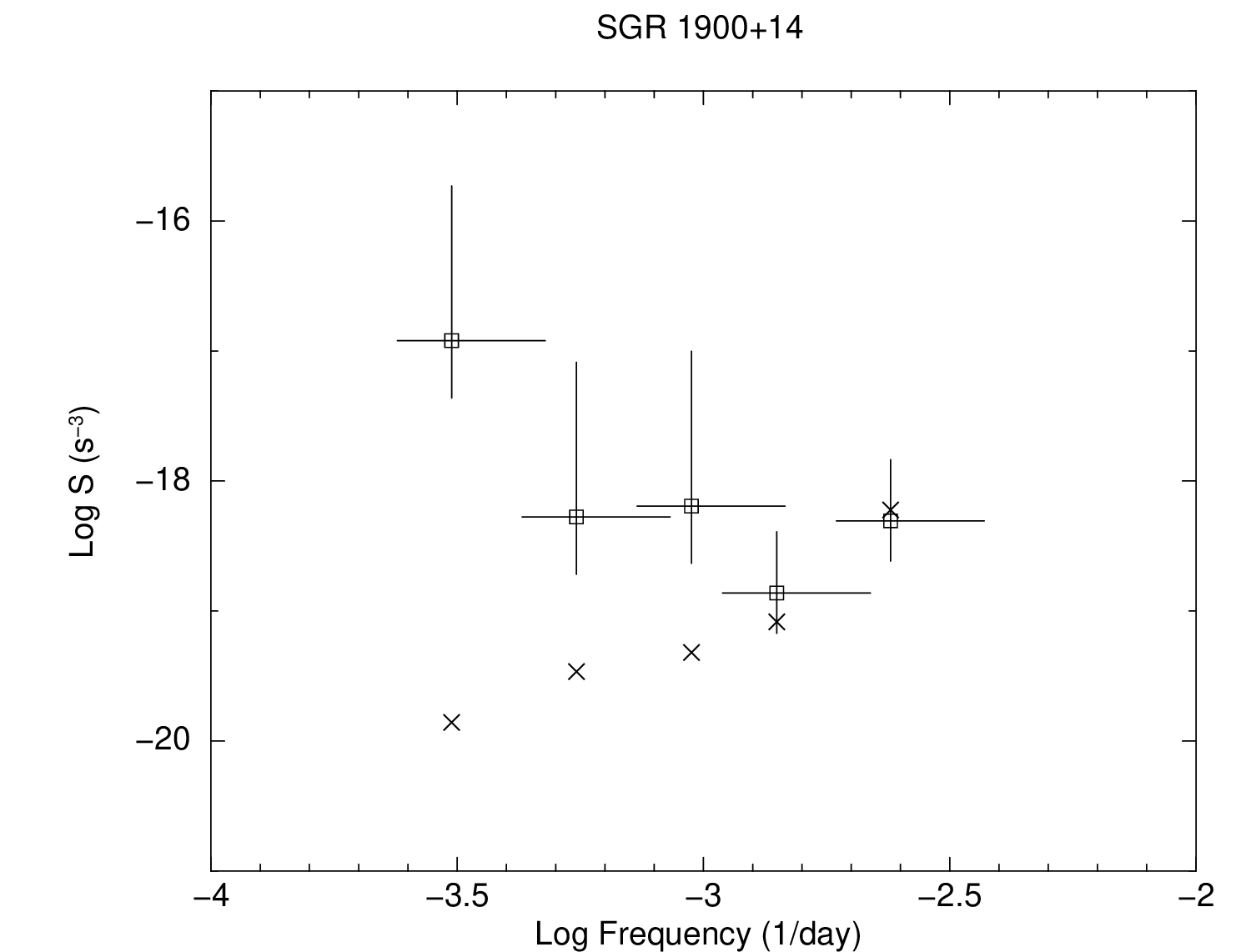} \hspace{0.5 cm}\includegraphics[width=5.4cm, angle=270]{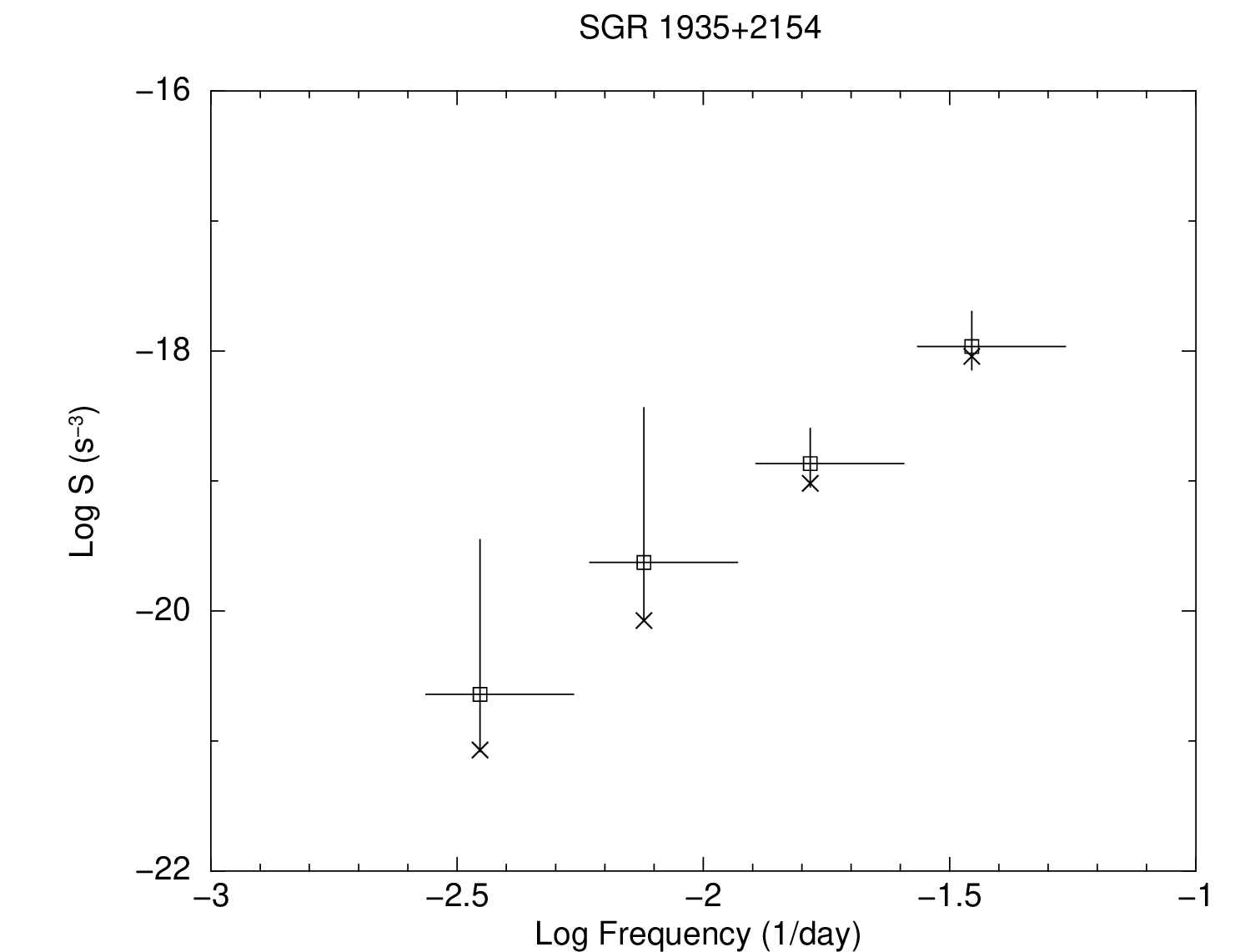} 
  
  \vspace{0.2 cm}
  
  \includegraphics[width=5.4cm, angle=270]{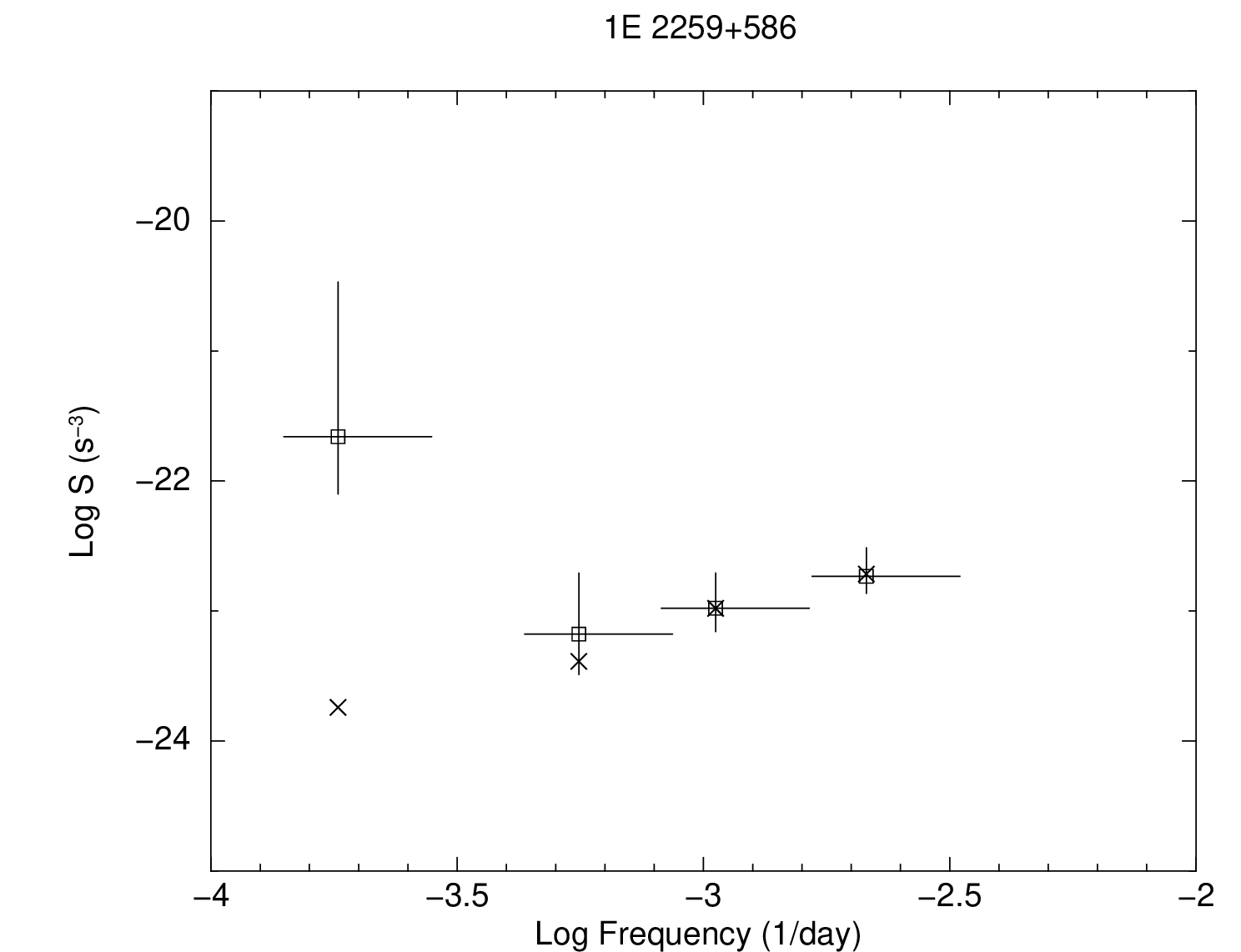}}
  \caption{continued.}
  \label{power}
\end{figure*}

Earlier power density spectra estimates of SGR 1900+14 and SGR 1806--20 have shown steep power density spectra on short time intervals with power law indices $-3.7\pm0.6$ and $-3.6\pm0.7$, respectively \citep{woods2002}. In this study, we construct the power density spectra of SGR 1900+14 and SGR 1806--20 using extended data sets. At shorter time scales, the steepness of the power density spectra are consistent with the values reported by \cite{woods2002} but at longer time scales the power spectra become whiter. Therefore, the overall power law indices are reduced to  $-2.61\pm1.49$ and $-2.58\pm0.46$ for SGR 1900+14 and SGR 1806--20, respectively. The power density spectra of SGR 1900+14 and SGR 1806-20 are rather complicated; the power law index changes on different analysis frequency ranges.

1E 1048.1--5937 shows red noise behaviour with power law index of $-1.00 \pm 0.22$ while
SGR J1745--2900 and 1E 1841--045 exhibit high level of timing noise at low frequencies implying a red noise behaviour at long time scales. However, power estimates at higher analysis frequencies for SGR J1745--2900 and 1E 1841--045 are in agreement with white noise model in pulse frequency derivatives. 
 
The power density spectra of 1E 2259+586, SGR 0418+5729,  SGR 0501+4516, 1RXS J170849.0--400910, SGR 1833--0832 and XTE J1810--197 track the measuremental noise levels except for only one or two high power estimate at low frequencies. The power density spectrum of SGR 1935+2154 follows the trend of measuremental noise levels. 

4U 0142+61, 1E 1547.0--5408 and CXO J164710.2--455216 have power law indices consistent with zero, so the pulse frequency derivative fluctuations are in accord with white noise. 
 
For the remaining eight magnetars (CXOU J010043.1--721134, CXOU J171405.7--381031, 3XMM J185246.6+003317, SGR 0526--66, Swift J1822.3--1606, Swift J1834.9--0846, SGR 1627--41 and PSR J1622--4950), the data samplings of spin frequency measurements are not adequate to construct power density spectra. Therefore, we only measure timing noise strength of each of these sources using existing data sets. 

Recent observations of magnetars  mostly take place during outburst activity. In most of the power spectra, we have seen red noise component at low frequencies which might be due to the effect of outbursts seen in these sources such as SGR J1745-2900 \citep{kaspi2014}, XTE J1810-197 \citep{camilo2016} and SGR 1806-20 \citep{woods2007, younes2015, younes2017}. Distribution of 
outbursts may determine the red noise component, i.e, single outburst may create 
excess low frequency noise, or a series of outbursts may create continous red noise 
process.  However, the sources 4U 0142+61 and 1E 1547.0-5408 exhibit white noise in power spectra despite the presence of their outbursts.  

1E 1547.0-5408 show only a marginal excess at the lowest frequency.
Following its 2009 outburst, the spin-down rate of 1E 1547.0-5408 approximately recovered back to its value at 2007 \citep{dib2012}. Therefore, an excess noise appeared only on the longest time scale (or the lowest frequency), and  $\dot{\nu}$ fluctuations at shorter time spans do not significantly alter the
shape of the power spectra which results in an absence of a red noise component at shorter time scales.

Bursting behaviour of 4U 0142+61 is similar to that of other magnetars. However, lack of a red noise component in
the power spectra of 4U 0142+61 might also be a consequence of  removal of the previously reported glitch
events (see \cite{archibald2017} and references therein) which are associated with outbursts, in our
analysis. Indeed, 4U 0142+61 exhibits quiet timing behavior between glitch events compared with other magnetars and its spin evolution is represented by low-order polynomials \citep{dib2014}.

\cite{baykal1999} investigated the stability of second derivatives of spin frequencies of four radio pulsars using rms residual technique. They suggested that the second derivatives of spin frequencies may originate from red noise processes arising from external torque fluctuations at the magnetosphere of the pulsars \citep{baykal1999}. The power spectra analyzed by \cite{baykal1999} exhibit red noise with power law indices varying between -0.39 and -2.41, however; they studied only a few systems with long time spans (in the order of 10000 days). On the other hand, these power spectra are observed to be flat (i.e. consistent with white noise) on shorter time scales in contrary to some magnetars showing red noise at similar time scales.

\subsection{Noise Correlations}

\begin{figure*}
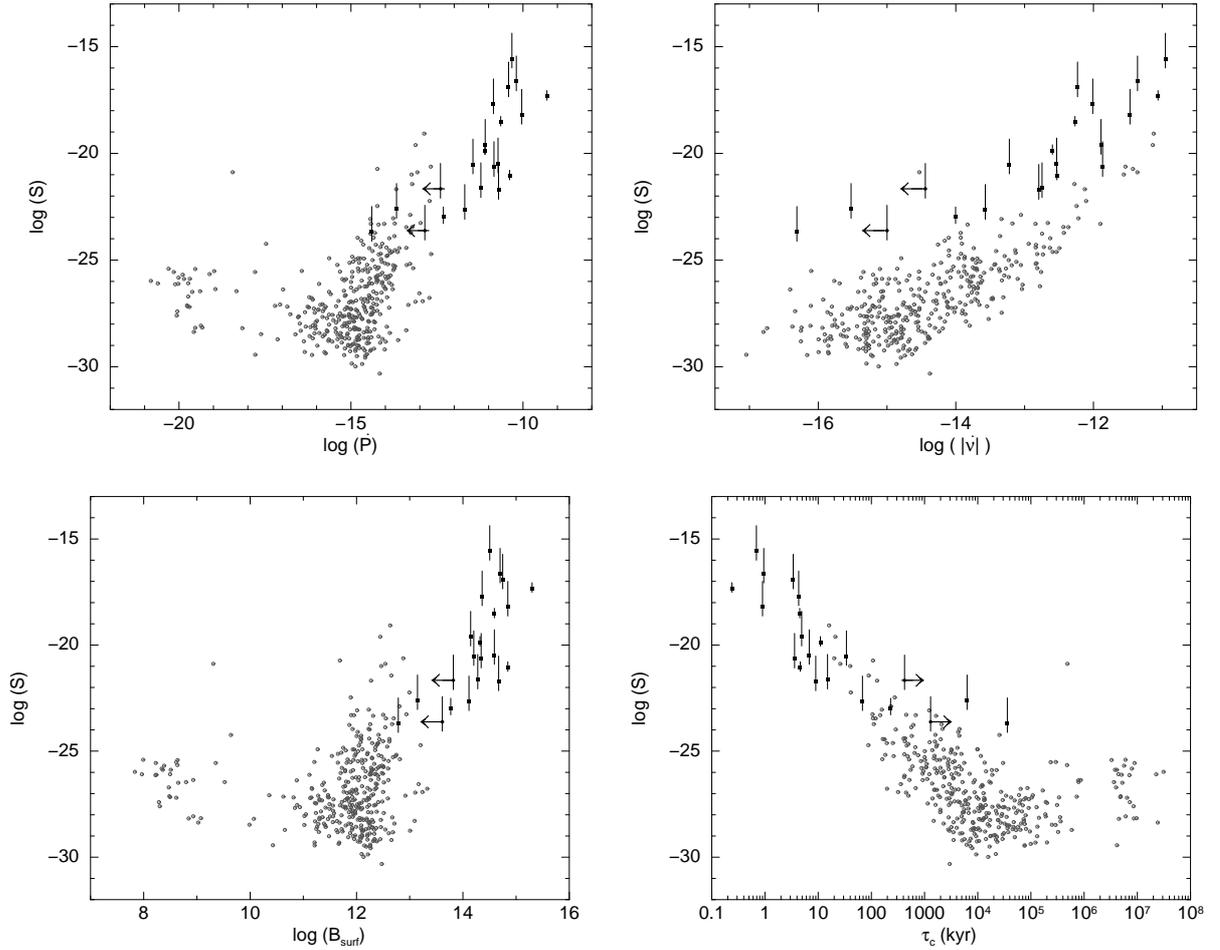

  \centering{
  \includegraphics[width=6cm, angle=270]{corr/pulsars_sd_pdot.eps} \includegraphics[width=6cm, angle=270]{corr/pulsars_sd_nu.eps}
  
  \vspace{0.5 cm}
  
  \includegraphics[width=6cm, angle=270]{corr/pulsars_b.eps} \includegraphics[width=6cm, angle=270]{corr/pulsars_age.eps}}
  \caption{Correlations between noise strength and period derivative (upper-left), frequency derivative (upper-right), surface magnetic field strength (lower-left), characteristic age (lower-right) for magnetars (filled black dots (this work)) and radio pulsars (empty gray dots \citep{hobbs2010}).}
  \label{corrp}
\end{figure*}

To compare the timing noise behaviour of magnetars with radio pulsars, we converted the reported rms values of 366 pulsars \citep{hobbs2010} to the noise strength $S$, using the corresponding normalization factor from Table 1 of \cite{deeter1984}. For magnetars, if available, we use the timing noise strength of approximately 1000 days time span, otherwise we use the noise strength corresponding to the maximum time span of the data set (see Table \ref{sources}). Then, we seek correlations between the noise strength values of magnetars and the other physical parameters such as $\dot{\nu}$, $\dot{P}$, $L_{x}$, and $B$ since such correlations are observed for pulsars \citep{cordes1980,baykal1993,hobbs2004,hobbs2010}. In figure \ref{corrp}, we demonstrate our results on magnetars together with the sample set of radio pulsars presented by \cite{hobbs2010}.

In figure \ref{corrp} upper-left panel, we present noise strength versus period derivative for the sample set of pulsars and magnetars. Our results indicate that there exist a correlation between noise strength and $\dot{P}$ for magnetars with Pearson correlation coefficient of $p=0.78$. \cite{cordeshelfand1980} studied the timing behavior of 50 pulsars and found that the timing noise of these objects are correlated with their period derivative and uncorrelated with radio luminosity. Our results on magnetars exhibit a correlation analogous to pulsars \citep{hobbs2010} and forms a continuum with the pulsar population. In figure \ref{corrp} upper-right panel, a similar kind of correlation can be observed between $S$ and  $\dot{\nu}$ as well ($p=0.85$, for magnetars only). However, the noise strengths of magnetars seem to be higher than those of the pulsars with same $\dot{\nu}$ possibly owing to either the increased timing activities during magnetar outbursts
\citep{dib2014} or the noise dominated $\ddot{\nu}$ values \citep{baykal1999}. For the first case, increased timing activities during outbursts may explain the higher timing noise
levels observed in magnetars, especially when we consider that neutron stars with higher initial
magnetic fields (B$> 10^{14}$G) tend to have more frequent bursting behaviors \citep{pons2011, pernapons2011}. For the latter case, if a pulsar spins down solely due to magnetic braking, second derivative of the spin frequency should be $\ddot{\nu}=\frac{n\dot{\nu}^2}{\nu}$ with the braking index $n=3$. But the observed values of braking indices in pulsars vastly differ from pure dipole braking value and vary between $-287\,\,986$ to $+36\,\,246 $ \citep{hobbs2010}. Therefore, it is suggested that the observed values of $\ddot{\nu}$ are not originated from pure dipole braking but they are dominated by timing noise \citep{baykal1999,hobbs2004,hobbs2010}.

In Figure \ref{corrp} lower-left panel, we illustrate the noise strengths of the sample set of pulsars and magnetars as a function of their inferred magnetic dipole field strength. \cite{tsang2013} studied the timing noise behaviour of several AXPs (1E 1841--045, RXS J170849.0--400910, 1E 2259.1+586, and 4U 0142+61) along with a large set of pulsars. They observed a correlation between the frequency noise and magnetic field which is attributed to the variations in the magnetospheric moment of inertia. We find a similar kind of correlation among magnetars with a Pearson correlation coefficient $p=0.71$.  According to magnetar model, a connection between timing noise and B-field is expected since most of the physical processes
that increase the level of the timing noise (i.e; decay rate and structural changes of magnetic field, crust craking, internal stresses and outbursts, etc.) are governed by the strength of magnetic field
\citep{thompson2002, pons2011, beloborodov2009}. We also find that there is an anti-correlation between characteristic age $\tau_c$ and the timing noise strength of the magnetars (see Fig. \ref{corrp} lower-right panel, $p=-0.82$) which is also observed for pulsars \citep{hobbs2010}. The anti-correlation between $\tau_{c}$ and the noise strength further supports the idea that the timing noise level is decreasing as the magnetic field of the source decays. Considering all the correlations, the timing noise strengths of
magnetars and pulsars are seem to be linked and they posses similar noise floor that is possibly associated with their magnetic fields as suggested by \cite{tsang2013}. A link between these two neutron star populations was also suggested considering their surface temperatures, quiescent X-ray luminosities, magnetic fields \citep{kaspi2010, an2012} and magnetar-like bursts observed from the high B-field
pulsars PSR J1119-6127 \citep{Gogus2016, archibald2017, archibald2018} and PSR J1846-0258 \citep{Gavriil2008}. Moreover, both the "magnetar-like pulsar" PSR J1846-0258 \citep{Ng2008} and  the magnetar Swift J1834.9-0846 \citep{younes2016} are sourrounded by pulsar wind nebulae.

\begin{figure}
  \centering{
  \includegraphics[width=6.8cm, angle=270]{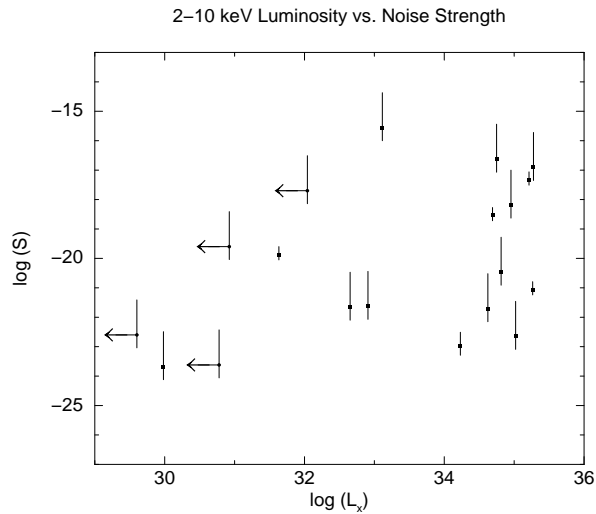}}
  \caption{X-ray Luminosity versus noise strength of magnetars.}
  \label{lum}
\end{figure}

\cite{baykal1993} studied a set of accreting pulsars, and reported a correlation between X-ray luminosity and timing noise strength. In accreting pulsars, mass transfer episodes enhance X-ray luminosity and exert external torque on the pulsar, implying that the noise strength of accreting pulsars grow with mass accretion rate. In Figure \ref{lum}, we present noise strength versus X-ray luminosity of magnetars in which we do not observe any correlation among these parameters ($p=0.41$). This implies that the fall-back disk model \citep{alpar2001}, which suggests that the observed X-ray emission and spin-down behaviour of magnetars are sustained by accretion from a debris disk after the supernova explosion, is a less likely explanation for the timing noise of these sources.
Therefore, the timing noise of magnetars possibly arise from fluctuations of high dipolar fields, rather than fluctuations of external torque exerted by a debris disk.
Together with the results stated above, we conclude that the timing noise behaviour of magnetars form a continuum with radio pulsar population, rather than accreting sources.

\section{Summary}

In this work, we for the first time present the extended timing noise analysis of all magnetars. We measure the noise strengths of all magnetars and construct the power density spectra of fifteen of them. We see that the noise strength is correlated with frequency derivative, magnetic field strength and anti-correlated with age. By comparing these correlations with those found for radio pulsars \citep{hobbs2010}, we observe that magnetars exhibit a similar timing noise behaviour with radio pulsars. However, magnetars do seem to have {\bf{1)}} higher timing noise levels compared with those of radio pulsars and {\bf{2)}} red noise components arising at shorter time scales which are expected as a consequence of the magnetar model \citep{thompsonduncan95, thompsonduncan96}, since a stronger initial magnetic field leads to increased chances of magnetic structure variations, more frequent bursting behavior and greater internal stresses \citep{pons2011, pernapons2011}. On the other hand, we do not see a correlation between noise strength of magnetars and their X-ray luminosity, which was observed for accreting sources \citep{baykal1993}. Therefore our findings put a further evidence that magnetar population are in continuum with radio pulsar population as suggested by \cite{kaspi2010,an2012,tsang2013}. Our results on noise correlations together with the power density spectra of magnetars, imply that the noise process in magnetars is associated with magnetospheric moment of inertia fluctuations \citep{tsang2013}. The noise strengths of magnetars follow the correlations that pulsars show with age and magnetic field. This may imply that the physical processes governing timing noise in both populations are similar as it is independently suggested for glitch events observed in both populations \citep{dib2014}. 

\section*{Acknowledgements}

We acknowledge support from T\"{U}B\.{I}TAK, the Scientific and Technological Research Council of Turkey through the research project MFAG 114F345. We are grateful to anonymous referee for useful comments.  A.B. acknowledges Prof. Werner Becker and Max Planck Institute for Extraterrestrial Physics (MPE) for invitation as a guest.  




\bibliographystyle{mnras}
\bibliography{magnetars_v2.0} 








\bsp	
\label{lastpage}
\end{document}